\documentclass[12pt]{article}
\usepackage{color}
\usepackage{bm}
\usepackage{slashed}
\usepackage{varioref}
\usepackage{xcolor} 
\usepackage{graphicx}
\usepackage{amsmath}
\usepackage{amssymb}
\usepackage{amsfonts}
\usepackage{amsthm}
\usepackage{array}
\usepackage{yfonts}
\usepackage{float}
\usepackage{comment}
\usepackage{caption}
\usepackage[sort]{cite}

\usepackage[
      colorlinks=true,
      linkcolor=blue,
      urlcolor=blue,
      filecolor=blue,
      citecolor=red,
      pdfstartview=FitV,
      pdftitle={},
      pdfauthor={},
      pdfsubject={},
      pdfkeywords={},
      bookmarksopen=true
]{hyperref}

\usepackage{epsfig}

\usepackage{hyperref}
\def\beq{\begin{eqnarray}}
\def\eeq{\end{eqnarray}}

\def\be{\begin{equation}}
\def\ee{\end{equation}}
\def\bea{\begin{eqnarray}}
\def\eea{\end{eqnarray}}

\def\be{\begin{equation}}
\def\ee{\end{equation}}
\def\bea{\begin{eqnarray}}
\def\eea{\end{eqnarray}}
\labelformat{equation}{Eq.~(#1)} 
\labelformat{appendix}{Appendix #1}

\def\nn{\nonumber}

\newcounter{mnotecount}[section]

\renewcommand{\themnotecount}{\thesection.\arabic{mnotecount}}

\newcommand{\mnotex}[1]
{\protect{\stepcounter{mnotecount}}$^{\mbox{\footnotesize
$
\bullet$\themnotecount}}$ \marginpar{
\raggedright\tiny\em
$\!\!\!\!\!\!\,\bullet$\themnotecount: #1} }

\date{}
\numberwithin{equation}{section}

\begin{document}

\title{\bf Gravitational multipole moments\\ for asymptotically de Sitter spacetimes}

\author{ Sumanta Chakraborty\footnote{tpsc@iacs.res.in}$~^{1}$, Sk Jahanur Hoque
\footnote{jahanur.hoque@utf.mff.cuni.cz}$~^{2}$, Roberto Oliveri\footnote{roliveri@fzu.cz}$~^{3}$
\vspace{0.6cm} \\
$^{1}${\small{School of Physical Sciences, Indian Association for the Cultivation of Science, Kolkata 70032, India}}
\vspace{0.3cm} \\
$^{2}${\small{Institute of Theoretical Physics,
Faculty of Mathematics and Physics, Charles University,}}
\\ 
{\small{V~Hole\v{s}ovi\v{c}k\'ach 2, 180~00 Prague 8, Czech Republic}}
\vspace{0.3cm} \\
$^{3}${\small{CEICO, Institute of Physics of the Czech Academy of Sciences,}}\\
{\small{Na Slovance 2, 182 21 Praha 8, Czech Republic}}
}
\maketitle 

\begin{abstract}

We provide a prescription to compute the gravitational multipole moments of compact objects for asymptotically de Sitter spacetimes. Our prescription builds upon a recent definition of the gravitational multipole moments in terms of Noether charges associated to specific vector fields, within the residual harmonic gauge, dubbed multipole symmetries. We first derive the multipole symmetries for spacetimes which are asymptotically de Sitter; we also show that these symmetry vector fields eliminate the non-propagating degrees of freedom from the linearized gravitational wave equation in a suitable gauge. We then apply our prescription to the Kerr-de Sitter black hole and compute its multipole structure. Our result recovers the Geroch-Hansen moments of the Kerr black hole in the limit of vanishing cosmological constant.
\end{abstract}

\newpage

\tableofcontents

\section{Introduction and summary of the results}

The multipole moments associated with a gravitational field have always been relevant and important in the study of various solutions arising from General Relativity, since its early days. These studies involving multipole moments have impacted many areas of research ranging from mathematical physics to astrophysics. The study of multipole moments in various contexts has become even more timely, since the discovery of gravitational waves \cite{Abbott:2020khf,Abbott:2020uma,LIGOScientific:2018mvr,Abbott:2017gyy,TheLIGOScientific:2017qsa,Abbott:2016blz,TheLIGOScientific:2016pea}.
The observation of gravitational waves from coalescence of binary compact objects can have potential applications in order to address questions about the nature of compact objects, such as black holes, neutron stars and the binary systems thereof \cite{Ryan:1995wh,Ryan:1997hg,Berti_2015,Cardoso:2016ryw,Barack:2018yly,Barausse:2020rsu,Maggio:2021ans}.
Future space-based gravitational wave detectors, besides mass and spin, will be also able to measure the quadrupole mass moment of a supermassive object in a binary system with great accuracy \cite{Ryan:1997hg,Barack_2007,Babak_2017}, 
as well as put bounds on four leading order multipole moments in the high mass ratio ($\sim 10$) 
limit \cite{Kastha_2019}.

In General Relativity, the gravitational field is decomposed in two sets of multipole moments --- the mass and the spin moments \cite{doi:10.1098,poisson_will_2014,Blanchet_2014}. Using the Penrose's conformal completion technique, a geometrical definition of the multipole moments for static, asymptotically flat spacetimes was pioneered by Geroch \cite{Geroch:1970cc,Geroch:1970cd}. Later, Geroch's definition was generalised to the stationary case by Hansen \cite{Hansen:1974zz}. Subsequently, Beig, Simon and Kundu \cite{BeigSimon1,BeigSimon2,Kundu1,Kundu2,Beig:1981zz} had developed further properties of the multipole moments for stationary and asymptotically flat spacetimes. On the other hand, imposing no incoming radiation for linearized radiating gravitational fields, Thorne \cite{RevModPhys.52.299} provided a definition of multipole moments for asymptotically flat spacetimes. Thorne's moments are defined within the harmonic gauge, upon further fixing the residual gauge, where the gravitational field is expanded in spherical harmonics and the multipole moments are read from such an expansion. Thorne's definition applies also to stationary non-linear configurations in General Relativity and it is shown to be equivalent to the Geroch--Hansen’s definition for stationary spacetimes, up to a choice of normalization \cite {1983GReGr..15..737G}. More recent developments about gravitational multipole moments in asymptotically flat spacetime can be found in \cite{Hoenselaers_1990,Sotiriou_2004,Fodor,Backdahl:2005be,B_ckdahl_2005, Backdahl:2006ed,Compere:2017wrj}.

Intriguingly, the multipole moments of a black hole have very simple structures: 
\emph{e.g.}, the multipole moments of a Kerr black hole only depends on its mass and spin \cite{Geroch:1970cd,Hansen:1974zz}. However, such is not the situation for neutron stars or for the exotic compact objects. The multipole structures of these objects are much more complex and are very much distinct from one another. Thus the study of the multipole structure of a compact object may reveal its true nature and interesting properties. For example, for a Kerr black hole all the odd mass moments and even spin moments are identically zero, however for certain fuzzball states odd mass moments and even spin moments do exist \cite{Bena:2020see,Bena:2020uup,Bianchi_2020,Bianchi_2021}. Therefore, if the gravitational wave observations tell us that the odd mass moments and even spin moments of the merging compact objects are indeed non-zero, then the fuzzball models will become particularly relevant. This shows the importance of understanding the multipole moments of compact objects in greater detail.    

As emphasized earlier, the multipole moments of gravitational fields are well understood for asymptotically flat spacetimes, however they remain largely unexplored for asymptotically non-flat and in particular for asymptotically de Sitter spacetimes. This is because the Geroch-Hansen as well as the formalism of Thorne depend crucially on the asymptotic flatness of the spacetime. Though there are instances where the formalism can be applied for asymptotically non-flat spacetimes: this is the case for spacetimes with NUT charge \cite{Mukherjee_2020}, because the relevant codimension-1 hypersurfaces are asymptotically flat even in the presence of NUT charge and the Geroch-Hansen formalism is readily applicable. On the other hand, for asymptotically de Sitter spacetimes these codimension-1 hypersurfaces are non-flat, rendering the known analysis of multipole moments inapplicable. 

The aim of this work is to start filling this gap in the literature and provide a definition of gravitational multipole moments for asymptotically de Sitter spacetimes. 

The mass and spin multipole moments have been recently defined for a generic metric theory of gravity in terms of the Noether charges associated with the so-called multipole symmetries,\footnote{The concept of multipole symmetries and the relation with multipole moments has been originally introduced in \cite{Seraj_2017} in the context of Maxwell electrodynamics.} that are specific residual transformations in the harmonic gauge \cite{Compere:2017wrj}. This definition also agrees with the previous results of Thorne for linearized radiating spacetimes and the Geroch-Hansen formalism for stationary spacetimes. At this outset, let us briefly summarize the main strategy to compute multipole moments in \cite{Compere:2017wrj}. The main idea is to extract the multipole moments from the radial expansion of the metric tensor. For asymptotically flat spacetimes, following Thorne \cite{RevModPhys.52.299}, one has to reach the harmonic gauge as explained earlier. Upon fixing the harmonic gauge, one is left with its residual gauge transformations. In \cite{Compere:2017wrj}, it was demonstrated that the multipole symmetries are actually residual gauge transformations preserving the asymptotic behaviour of the lapse function and shift vector. The next step in the analysis of \cite{Compere:2017wrj} is to compute the Noether charges associated with the multipole symmetries. Explicit expressions for the surface charges in the four-dimensional General Relativity can be found in \cite{Abbott:1981ff,Barnich:2001jy}. The multipole moments of the solution are identified from the Noether charges upon a regularization procedure.

In this paper, we wish to extend the method of \cite{Compere:2017wrj}, as outlined above, for asymptotically de Sitter spacetimes. More specifically, we achieve three main results --- (a) Our first result is to write down the de Sitter spacetime in harmonic gauge; surprisingly, de Sitter spacetime was never written in the harmonic gauge before, to the best of our knowledge. Expressing the de Sitter spacetime in the harmonic gauge is instrumental to derive the residual harmonic gauge transformations and, among these, the multipole symmetries preserving certain fall-off conditions for asymptotically de Sitter spacetimes. These fall-off conditions are such that they preserve the lapse function and the shift vector, analogously as in \cite{Compere:2017wrj}. (b) We also provide an alternative derivation of the multipole symmetries to be those residual gauge transformations that eliminate the non-propagating degrees of freedom in the linearized outgoing wave solutions around de Sitter background. This different interpretation of multipole symmetries sheds more light over the physics of the multipole symmetries and complements the original derivation carried out in \cite{Compere:2017wrj}. (c) Finally, in order to apply our findings on a concrete example, we compute the multipole structure of the Kerr-de Sitter black hole. In particular, we provide explicit computation of the first few mass and spin multipole moments. Our expressions reproduce the well-known mass and angular momentum of the Kerr-de Sitter black hole and, in the limit of asymptotically flat spacetime, we recover the Geroch-Hansen's multipole moments of the Kerr black hole.

The paper is organized as follows. In section \ref{dS}, we briefly present different coordinate systems for the de Sitter spacetime, which will be relevant for our purpose. Section \ref{Harmonic} deals with writing down the de Sitter metric in the harmonic gauge. In section \ref{Sec.4}, exploiting the residual harmonic gauge freedom, we obtain the multipole symmetries for asymptotically de Sitter spacetimes. In section \ref{sec:5}, we provide the linearized gravitational wave equation in de Sitter background, and we obtain the multipole symmetries from the residual gauge transformations of the linearized theory. Finally, in section \ref{Sec:KdS}, we compute the multipole moments of the Kerr de Sitter spacetime. We conclude with a summary and perspectives for future works. Two appendices contain --- (a) the technical derivation of converting the de Sitter metric to harmonic coordinates and (b) the exact analytic expressions of the first few mass and spin multipole moments of the Kerr-de Sitter black hole.

\bigskip

\emph{Notation and conventions:} We adopt the mostly positive signature convention, \emph{i.e.}, the Minkowski metric in the Cartesian coordinate system takes the form: $\textrm{diag}(-,+,+,+)$. The Greek indices run over all the spacetime coordinates, while the Latin indices run over the spatial coordinates. Furthermore, we set the fundamental constants, such that $c=1=G$. 

\section{Brief review of the de Sitter spacetime} \label{dS}

We briefly review the de Sitter spacetime to point out the key features and the coordinate charts we will be using in this work. Unlike the Minkowski spacetime, which admits a natural, global Cartesian coordinate chart, the de Sitter spacetime has several charts appropriate for different situations --- the global, Poincar\'e and static patches. Furthermore, the de Sitter spacetime is a solution to the equations: $R_{\mu \nu}=\Lambda g_{\mu \nu}$, where $\Lambda$ is referred to as the cosmological constant term.

\begin{figure}[!ht]
\centering
\begin{center}
\includegraphics[scale=0.5]{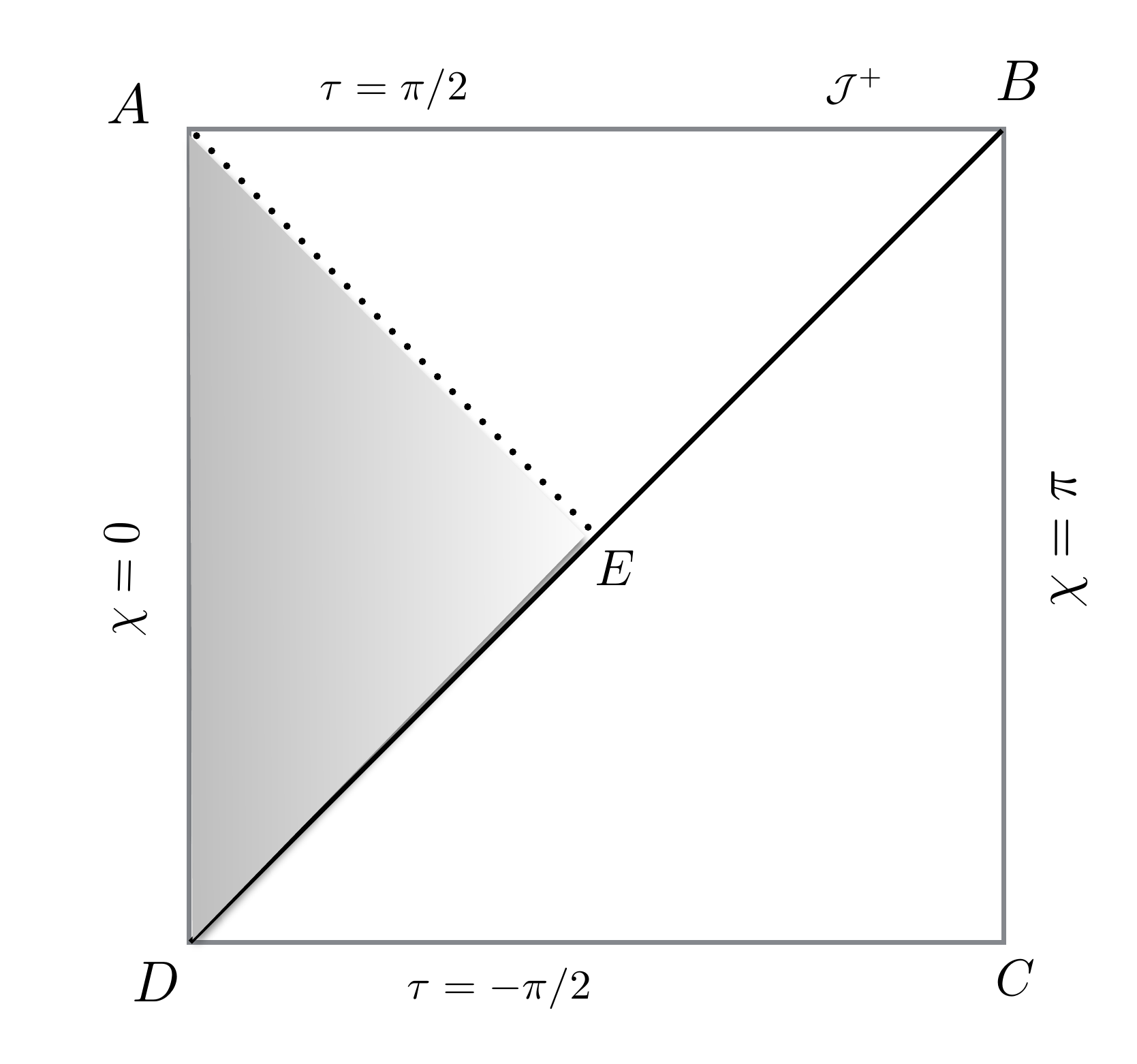} 
\caption{ABCD denotes the \emph{global chart}, ABD is a \emph{Poincar\'e patch}, while
AED is a \emph{static patch}.  The angular coordinates, $\theta$ and $\phi$, are
suppressed. An observer, represented by the world line
DA, has its causal future $J^+$ spanning the region DBA and is one of
the {\em Poincar\'e patches}. Its spacelike boundary, denoted by the
line AB, is the {\em future ``null'' infinity}, ${\cal J}^+$.
\label{GlobalChartFig}
}
\end{center}
\end{figure}

The de Sitter spacetime is most naturally defined as the hyperboloid in the five dimensional Minkowski spacetime, which also introduces a set of coordinates covering the full de Sitter spacetime, known as the {\em global patch}. The global patch is covered by the coordinates $(\tau, \chi, \theta, \phi)$, and the de Sitter metric is given by
\be\label{dS_global}
ds^{2}=-d\tau^{2}+\frac{1}{H^{2}}\cosh^{2}(H\tau)\left[d\chi^2 +\mathrm{sin}^2\chi \left(d\theta^2 + \mathrm{sin}^2\theta d\phi^2\right)\right]~,
\ee
where $H:=\sqrt{(\Lambda/3)}$ is the Hubble constant. As evident from the above metric, the global topology of the de Sitter spacetime is $\mathbb{R}\times S^{3}$; see Fig.~\ref{GlobalChartFig}. In the global chart, the axial symmetry of the de Sitter spacetime is guaranteed by the Killing vector $(\partial/\partial \phi)$. However, unlike the Minkowski spacetime, $(\partial/\partial \tau)$ is not a Killing vector in the global coordinates. 

Among the other coordinate systems associated with the de Sitter spacetime, the \emph{Poincar\'e patch}, which constitutes the causal future (past) of observers and covers ``half'' of the global chart is of much interest. As we will see, in the cosmological context and for our current purpose of studying compact sources in the de Sitter background, the Poincar\'e patch will turn out to be very useful. There are two natural coordinate charts for the Poincar\'e patch: (a) the {\em conformal chart} with coordinates $(\eta, x^i)$, and (b) the {\em cosmological chart} with coordinates $(t, x^i)$. In the conformal chart ($\eta, x^i)$, the de Sitter metric takes the form
\begin{eqnarray}\label{ConformallyFlat}
ds^{2}=a^{2}(\eta)~\left(-d\eta^{2}+\delta_{ij}dx^idx^{j}\right)~,
\end{eqnarray}
where $a(\eta)=-(H\eta)^{-1}$ with $\eta \in (-\infty,0)$. A drawback of the conformal coordinates is that they are not suitable for studying the flat limit $\Lambda \rightarrow 0$ (or, equivalently $H\rightarrow 0$); see, \emph{e.g.}, \cite{ABKIII} (however also see, \cite{Rajeev:2019okd}). Thus one is led to adopt the cosmological coordinates, with the cosmic time $t$ being related to the conformal time $\eta$ via $\eta:=-(1/H)e^{-Ht}$. In these coordinates ($t,x^{i}$), the line element for the de Sitter spacetime becomes
\begin{eqnarray}\label{PlanarMetric}
ds^2=-dt^2+e^{2Ht}\left(\delta_{ij}dx^idx^{j}\right)~.
\end{eqnarray}
This is the most well-known form for the de Sitter metric and it will be used extensively in this work. 

The Poincar\'e patch has a seven dimensional isometry group --- \emph{three} spatial rotations, \emph{three} spatial translations and \emph{one} time translation. In order to find out the Killing vector field of time translations, we let $t\rightarrow t+\delta t$ in \ref{PlanarMetric}, where $\delta t$ is taken to be infinitesimal. This time translation changes the de Sitter line element and hence in order to make the de Sitter metric in the Poincar\'e patch invariant, we must have the following transformation for the spatial coordinates: $x^{i}\rightarrow x^{i}-H x^{i}\delta t$. Hence, the Killing vector field generating the time translation in the cosmological coordinate system is
\begin{equation}\label{TimeTrans}
t^{\mu}\partial_\mu= \partial_{t}-Hx^i\partial_i~.
\end{equation}
This vector will also play an important role in the study of multipole symmetries of de Sitter spacetime. 

The third patch, corresponding to ``half'' of the Poincar\'e patch, is referred to as the {\em static patch}; see Fig.~\ref{GlobalChartFig}. This is a natural patch for an isolated body or a black hole with a stationary neighbourhood. This patch of de Sitter spacetime can be covered by static coordinates $(T,R,\theta,\phi)$. In these coordinates, the line element of the de Sitter spacetime becomes
\be\label{dS_static} 
ds^{2}=-\big(1-H^{2}R^{2}\big) dT^{2} +\frac{dR^2}{\big(1-H^{2}R^{2}\big)}+R^2\left(d \theta^2+\sin^{2}\theta d\phi^{2}\right).
\ee
The existence of two Killing vector fields, $(\partial/\partial T)$ and $(\partial/\partial \phi)$, is clear from the line element above; correspondingly, the spacetime has both axial and time-translational symmetries. It should be emphasized that the Killing\footnote{It is immediate to show that the Killing vector, defined in \ref{TimeTrans} in the cosmological coordinates, transforms to $(\partial/\partial T)$ in the static coordinates. For that purpose one simply has to note that $\partial/\partial t = (1-H^2R^2)(\partial/\partial T) + HR(\partial/\partial R)$ and $r(\partial/\partial r) = R (\partial/\partial R) + H R^2/(1-H^2R^2) (\partial/\partial T)$, such that, $(\partial/\partial t)-Hr(\partial/\partial r)=(\partial/\partial T)$.} vector field $(\partial/\partial T)$ becomes null on the cosmological horizon of the de Sitter spacetime located at $R=H^{-1}$.

Let us conclude this section with a side remark about the global isometry group in the three patches of the de Sitter spacetime. The \emph{ten} dimensional isometry group of the full de Sitter spacetime is reduced to a \emph{seven} dimensional subgroup in the Poincar\'e patch, and to a \emph{four} dimensional subgroup in the static patch \cite{ABKI}. The symmetry reduction of the global isometry group can be best understood along the following lines: the null hyperplane BD (see Fig.~\ref{GlobalChartFig}) can be thought of as adding an additional boundary to the full de Sitter spacetime. Hence, the symmetry generators which are not tangential to BD are absent in the Poincar\'e patch. Similarly for the static patch, the Killing fields which are not tangential to the cosmological horizon are not symmetry generators in the static patch.

\section{de Sitter metric in the harmonic gauge}\label{Harmonic}

The harmonic gauge, also known as the de Donder gauge, plays a crucial role in solving the Einstein's equations \cite{Blanchet_2014}. The harmonic gauge is usually the favourite gauge where to read off the multipole moments from the spherical harmonic decomposition of the metric tensor \cite{RevModPhys.52.299}. In addition, the residual harmonic gauge symmetry is intimately connected with the nature of the multipole moments of any compact object \cite{Compere:2017wrj}. 
Given the metric tensor $g_{\mu \nu}$, the harmonic gauge condition is given by 
\bea
\partial_{\nu}(\sqrt{-g}g^{\mu \nu})=0~.
\eea
To impose the harmonic gauge condition on the metric $g_{\mu \nu}$, we first perform the coordinate transformation $x '^{\mu}=f^{\mu}(x)$. Subsequently, imposing the harmonic gauge condition on the metric in the $x'^{\mu}$ coordinate system amounts to 
\bea
\partial_{\nu'}\left(\sqrt{-g'}g^{\mu' \nu'}\right)=\sqrt{-g'}\ \square_{g'} x'^{\mu}=0~,
\eea
where $\square_{g'}:= g^{\mu\nu} \nabla_{\mu}\nabla_{\nu}$. Hence, choosing \emph{each} coordinate $x'^{\mu}$ to be harmonic in the metric $g_{\mu \nu}$, \emph{i.e.}, satisfying $\square_{g'}x'^{\mu}=0$, ensures that the harmonic gauge condition holds in the new coordinates. We will employ this procedure to transform the de Sitter metric to harmonic coordinates in all the three coordinate patches discussed in the previous section.  

Intriguingly, despite being one of the most familiar solution to the Einstein's equations with a cosmological constant, to our knowledge, harmonic coordinates for de Sitter spacetime do \emph{not} exist in the literature. Even though we can provide the transformation of all the three coordinate patches to harmonic coordinates, here we provide the explicit coordinate transformation from the cosmological coordinates to the harmonic coordinates, since it will be of relevance in what follows. We relegate to Appendix \ref{App:harm} the explicit coordinate transformations that bring the de Sitter metric in global and static coordinates to the harmonic form.

As discussed in the previous section, the Poincar\'e patch of the de Sitter spacetime can be described by the cosmological coordinates $(t, x^i)$, with the associated line element given by \ref{PlanarMetric}, for which
\begin{equation}
\sqrt{-g}g^{\mu \nu} = \mbox{diag}\left(-a(t)^{3}, a(t), a(t),a(t)\right),
\end{equation}
where $a(t):=e^{Ht}$. Therefore, it follows that the cosmological coordinates are not harmonic since  $\partial_{\mu}(\sqrt{-g}g^{\mu 0})=-3H e^{3Ht}\neq 0$. In order to obtain the harmonic coordinates for de Sitter in the Poincar\'{e} patch, we introduce a new set of coordinates $\bar{x}^{\alpha}$, such that $\square_{g}\bar{x}^{\alpha}=0$, where $\square_{g}$ is associated with the metric given in \ref{PlanarMetric}. Expanding out the above harmonic gauge condition, we obtain the following differential equation for the new coordinates
\bea
\left[-\frac{1}{a^{3}}\partial_{t}\left(a^{3}\partial_{t}\right)+\delta^{ij}\frac{1}{a^{3}}\partial_{i}\left(a\partial_{j}\right)\right]\bar{x}^{\alpha}=0~.
\label{9XII20.3}
\eea
We now have to choose an appropriate coordinate transformation, such that the above differential equation can be satisfied. For our purpose it will suffice to perform the following coordinate transformation
\bea
\bar{x}=x,\quad \bar{y}=y,\quad \bar{z}=z,\quad \bar{t}=f(t)~,
\eea
where $f(t)$ is an arbitrary function of the cosmic time. It is straightforward to verify that \ref{9XII20.3} is indeed obeyed for all the spatial coordinates, while for the time coordinate it yields the following differential equation
\bea 
\left(\frac{d^{2}f}{dt^{2}}\right)+3H\bigg(\frac{df}{dt}\bigg)=0~.
\eea
Solving this differential equation for the function $f(t)$, equipped with the boundary conditions $f(t=0)=0$ and $(df/dt)(t=0)=1$ (these boundary conditions are necessary to have a smooth flat spacetime limit), we obtain,
\bea
\bar{t}:=f(t)=\frac{1}{3H}\big(1-e^{-3Ht}\big)~.
\eea
Therefore, the line element of the Poincar\'{e} patch of the de Sitter spacetime takes the following form in the harmonic coordinates
\bea\label{dS_Cosmo_Harmonic}
ds^{2}=-\frac{d\bar{t}^{2}}{\left(1-3H\bar{t}\right)^{2}} + \left(1-3H\bar{t}\right)^{-2/3}\left(dx^{2}+dy^{2}+dz^{2}\right).
\eea
It can again be verified that, with somewhat long but straightforward algebra, that the de Sitter metric in the above coordinate system indeed satisfies the harmonic gauge condition.

Having determined the harmonic coordinates for the de Sitter spacetime in the Poincar\'{e} patch, we will proceed to find out the residual gauge symmetry and the vector fields generating these symmetries. As we will demonstrate, these vector fields will be crucial in obtaining the multipole symmetries and hence the multipole moments of compact objects living in an asymptotically de Sitter spacetime. We will determine these multipole symmetry vector fields in the next section.  


\section{Residual harmonic gauge and multipole symmetries}\label{Sec.4}

We are now in the position to discuss the residual harmonic gauge transformations and hence derive the vector fields from which the multipole moments, in terms of the Noether charges, can be associated. 
We choose an operational definition of asymptotically de Sitter spacetimes such that in harmonic gauge our metric matches with that of \ref{PlanarMetric} in the asymptotic regime.
For our purpose it will suffice to consider a constant cosmological time $t$ such that the asymptotic regime corresponds to the $r\rightarrow \infty$ limit, which is also identical to the $r_{\rm phys}:= re^{Ht}\to \infty$. Since this corresponds to the point $B$ in Fig. \ref{GlobalChartFig}, which is the analogue to spatial infinity $i^{0}$ in asymptotically flat spacetime, we will use this limit to define the multipole moments. 
Therefore, following \cite{Compere:2017wrj}, here also we demand the following asymptotic conditions on the metric components 
\be \label{25XII20.1}
g_{0\mu}=\bar{g}_{0\mu}+\mathcal{O}(1/r_{\rm phys})
\ee
Here $\bar{g}_{\mu \nu}$ denotes the background de Sitter spacetime. It should also be noticed that \ref{25XII20.1} is a non-tensorial relation and hence must be used in the harmonic coordinates, such that $\bar{g}_{\mu \nu}$ satisfies the harmonic gauge condition (note that the radial distance remains the same in both cosmological and harmonic coordinates). The vector field $\xi^{\mu}$ will preserve this asymptotic conditions given in \ref{25XII20.1} if it satisfies \emph{asymptotically} the following relation 
\bea \label{25XII20.2}
\pounds_{\xi} \bar{g}_{0\mu}:= \xi^{\alpha}\partial_{\alpha} \bar{g}_{0\mu}+\bar{g}_{\mu\alpha} \partial_{0}\xi^{\alpha}+\bar{g}_{\alpha 0} \partial_{\mu} \xi^{\alpha} =\mathcal{O}(1/r_{\rm phys})~.
\eea
For the harmonic coordinates in the Poincar\'{e} patch and for the component $\mu =0$, \ref{25XII20.2} becomes $\xi^{\alpha} \partial_{\alpha} g_{00}+2g_{0\alpha} \partial_{0} \xi^{\alpha}=0$. This yields the following solution for the time component of the vector field $\xi^{\mu}$ in harmonic coordinates
\bea  \label{27XII20.1} \label{19V21.2}
\xi^{0}=\left(1-3H \bar{t}\right) \epsilon(\bm{x})+\mathcal{O}(1/r_{\rm phys})~,
\eea
where $\epsilon(\bm{x})$ is an arbitrary function of the spatial coordinates $\bm{x}$. Similarly, for the spatial components of the asymptotic condition given in \ref{25XII20.2}, we obtain the following differential equation for the vector field $\xi^{\alpha}$
\bea \label{19V21.1}
\xi^{\alpha} \partial_{\alpha} g_{0i}+g_{0\alpha} 
\partial_{i}\xi^{\alpha}+g_{i\alpha} \partial_{0} \xi^{\alpha}=\mathcal{O}(1/r_{\rm phys})~,
\eea
Using \ref{19V21.2} and \ref{19V21.1}, the solution for the spatial components, in the harmonic coordinates, is given by
\bea \label{killing_spatial}
\xi^{i}=\frac{1}{2H}\left[1-\left(1-3H\bar{t}\right)^{2/3}\right] \delta^{ij} \partial_{j} \epsilon(\bm{x}) + \zeta^{i}(\bm{x})+\mathcal{O}(1/r_{\rm phys})~.
\eea
Thus, it follows that under the diffeomorphism $x^{\mu}\rightarrow x^{\mu}+\xi^{\mu}$, the change in the de Sitter metric satisfies the necessary asymptotic boundary condition given by \ref{25XII20.1}, provided the time and spatial components of $\xi^{\mu}$ are given by \ref{27XII20.1} and \ref{killing_spatial}, respectively. 

The unknown functions, $\epsilon(\bm{x})$ and $\zeta^{i}(\bm{x})$, appearing in the components of the vector field $\xi^{\mu}$ above, are arbitrary as of now.
However the vector field should satisfy one more additional condition, namely the harmonic gauge condition $\square_{\bar{g}}\xi^{\mu}=0$ in the background de Sitter spacetime. We must emphasize that $\square_{\bar{g}}\xi^{\mu}=0$ should be understood as four scalar equations, for each one of the four functions $\xi^{\mu}$ (for a detailed discussion along these lines, see \cite{poisson_will_2014}). The time component of the above equation, $\square_{\bar{g}}\xi^{\alpha}=0$, demands that the function $\epsilon(\bm{x})$ must satisfy the following differential equation\footnote{As we are interested in computing multipole symmetry vector in asymptotic regime, we neglect $\mathcal{O}(1/r_{\rm phys})$ in the rest of the section to avoid cluttering in notation.}
\bea \label{27XII20.2}
\delta^{ij} \partial_{i} \partial_{j} \epsilon(\bm{x})=0~.
\eea
While the spatial components of the equation, $\square_{\bar{g}} \xi^{\alpha}=0$, yields the following differential equation for $\xi^{i}$, in the harmonic coordinates

\bea  \label{27XII20.3}
-\partial_{\bar{t}}^{2} \xi^{i}+\left(1-3H \bar{t}\right)^{-4/3} {} \delta^{kl} \partial_{k} \partial_{l} \xi^{i}=0~.
\eea
Substituting for the spatial components of the vector field $\xi^{i}$ from \ref{killing_spatial} and using \ref{27XII20.2}, we obtain the following differential equation for the spatial vector field $\zeta^{i}$
\bea\label{vec_zeta}
\delta^{kl} \partial_{k} \partial_{l} \zeta^{i}(\bm{x}) - H\delta^{ij} \partial_{j} \epsilon(\bm{x})=0~,
\eea
whose solution can be written as 
\bea\label{sol_zeta}
{\bm \zeta}= -{\bm r} \times \boldsymbol{\nabla}\epsilon_{1}(\bm{x})+\boldsymbol{\nabla}\epsilon_{2}(\bm{x}) - H{\bm r} + {\bm V},
\eea
where $\epsilon_{1}(\bm{x})$ and $\epsilon_{2}(\bm{x})$ are two harmonic functions satisfying \ref{27XII20.2}. Moreover, we have defined $\boldsymbol{\nabla}\epsilon:=\delta^{ij}\partial_{i} \epsilon(\bm{x})\partial_{j}$, and $\times$ denotes the cross product defined in the three 
dimensional Euclidean space. Further, the vector field ${\bm V}$ is the inhomogeneous part of  \ref{vec_zeta} satisfied by $\zeta^{i}$. This vector field is discarded since it does not play any role in General Relativity \cite{Compere:2017wrj}.

A similar analysis can be performed for the global and the static patch of the de Sitter spacetime as well. However, in our subsequent discussions, we will restrict ourselves to the harmonic coordinates of the de Sitter metric in the Poincar\'e patch, since this will turn out to be the most useful for the ensuing discussions. For this reason, we wish to express the above vector field $\xi^{\mu}$ in the cosmological coordinates, rather than in the harmonic coordinates. A simple coordinate transformation, following section~\ref{Harmonic}, between the two sets of coordinates, yields
\bea \label{killing_FRW}
\xi^{\mu}=\epsilon(\bm{x})\left(\partial_t\right)^{\mu}+\left[\frac{1}{2H}\left(1-e^{-2Ht}\right)\delta^{ij} \partial_{j}\epsilon(\bm{x})+\zeta^{i}(\bm{x})\right]\left(\partial_i\right)^{\mu}~.
\eea
Substituting the above solution for $\zeta^{i}$, as in \ref{sol_zeta}, in the above expression for the residual gauge vector field $\xi^{\mu}$ in the cosmological coordinates, we obtain,
\begin{align}\label{killing_FRW2}
\xi^{\mu}&=\epsilon(\bm{x})\left(\partial_t\right)^{\mu} 
\nonumber
\\ 
&\hskip 1 cm +\left[\frac{1}{2H}\left(1-e^{-2Ht}\right)\delta^{ij} \partial_{j}\epsilon(\bm{x})+\left(-\bm{r}\times \boldsymbol{\nabla}\epsilon(\bm{x})+\boldsymbol{\nabla}\epsilon(\bm{x})-H\bm{r}\right)^{i}\right]\left(\partial_i\right)^{\mu}~,
\end{align}
where $\epsilon(\bm{x})$ satisfies the equation $\nabla^{2}\epsilon(\bm{x})=0$, with $\nabla^{2}$ being the three-dimensional Laplacian operator. This vector field $\xi^{\mu}$ is the multipole symmetry vector field. In particular, we can decompose the above multipole symmetry vector field into three sets, namely
\begin{subequations}\label{MultipoleSymm}
\begin{align}
\label{MultipoleSymmK}
{\bm K_{\epsilon}}&:=\epsilon(\bm{x})\partial_{t}+\frac{1}{2H}\left(1-e^{-2Ht}\right){\bm\nabla} \epsilon(\bm{x})
-Hx^i \partial_i~, 
\\ 
\label{MultipoleSymmL}
{\bm L_{\epsilon}}&:=-{\bm r}\times {\bm \nabla} \epsilon(\bm{x})~, 
\\
{\bm P_{\epsilon}}&:={\bm \nabla} \epsilon(\bm{x})~.
\end{align}
\end{subequations}
Most importantly, in the limit, $H\rightarrow 0$, the above multipole symmetry vectors reduce to those of the asymptotically flat spacetime \cite{Compere:2017wrj}. Hence, following the flat spacetime analogy, one can identify the vector field $\bm{K}_{\epsilon}$ as the generator of the mass multipole moments, $\bm{L}_{\epsilon}$ as the generator of the spin multipole moments and $\bm{P}_{\epsilon}$ as the generator of the momentum multipole moments. Notice that, except for $\bm{K}_{\epsilon}$, both $\bm{L}_{\epsilon}$ and $\bm{P}_{\epsilon}$ are identical to their flat spacetime counterparts. This is expected, since the isometry group of the Poincar\'e patch includes both rotation and spatial translation symmetries as that of the flat spacetime, but the mass multipole symmetry vector gets modified by the presence of the cosmological constant. 

As evident from the above structure of the vector fields $\bm{K}_{\epsilon}$, $\bm{L}_{\epsilon}$ and $\bm{P}_{\epsilon}$, the multipole symmetries \ref{MultipoleSymm} depend on the harmonic function $\epsilon(\bm{x})$, whose decomposition consists of irregular and regular solid spherical harmonics. These are given by, $r^{-(\ell+1)}Y_{\ell m}(\theta,\phi)$ and $r^{\ell}Y_{\ell m}(\theta,\phi)$, respectively. The first branch, which are irregular at $r=0$, are simply gauge transformations and hence discarded\footnote{Another reason to discard the branch of irregular solid spherical harmonics is the following \cite{Compere:2017wrj}: we want to probe the multipole moments, that naively are the coefficients of the $1/r$ expansion of the metric tensor. In order to extract such coefficients, one needs a vector field that -- after being contracted with the metric tensor and derivative thereof -- gives us access to the $r^{-l}$ component of the metric tensor. This is achieved by the regular branch of the solid spherical harmonics.}; the second branch, instead, is used to decompose the harmonic function $\epsilon(\bm{x})$ as
\bea 
\epsilon(\bm{x})=\sum_{\ell=0}^{\infty}\sum_{m=-\ell}^{\ell}\epsilon_{\ell m}r^{\ell}Y_{\ell m}(\theta,\phi)~,
\eea
where $\epsilon_{\ell m}$ are arbitrary coefficients. We notice that for the $l=0$ and $l=1$ modes, the multipole symmetries in \ref{MultipoleSymm} reduce to the background symmetries of the de Sitter spacetime, discussed in section \ref{dS}. We will now demonstrate, that the vector associated with the residual gauge symmetry, also arises from the gauge freedom of the linear gravitational perturbations around de Sitter background. 

\section{Linearized perturbation of de Sitter spacetime}\label{sec:5}

In this section, we will consider linear gravitational perturbation of the de Sitter background in the cosmological coordinates. This will enable us to provide the perturbation equations in the cosmological coordinates along with the associated gauge choice simplifying the equations. It turns out that there is still a residual gauge freedom left in these perturbation equations, which enables us to eliminate the non-dynamical degrees of freedom and yields a symmetry vector identical to the one derived in section \ref{Sec.4}. This provides a completely independent way of deriving the multipole symmetries, further bolstering our claims in the previous section. We start by fixing the gauge condition associated with the linear perturbation equations.   

\subsection{Fixing the wave gauge: evolution of the linear gravitational perturbations}\label{Linear}

The gravitational perturbation around the de Sitter background is often considered in the conformal coordinates $(\eta,x,y,z)$ \cite{deVega:1998ia,DHI,ABKII}. However, for our current purpose, it will prove useful to consider the gravitational perturbations around de Sitter spacetime in the cosmological coordinates $(t,x,y,z)$. This is primarily because the residual gauge symmetry has the cleanest expression in the cosmological coordinates and can be compared to the corresponding expression for the flat spacetime with ease. The de Sitter metric in the cosmological coordinates has already been presented in \ref{PlanarMetric}, for which the non-zero components of the Christoffel connections are
\bea
\Gamma^{0} {}_{ij}= H e^{2Ht} \delta_{ij}~, \qquad
\Gamma^{i}{}_{0j}= H \delta^{i}_{j}~.
\eea
Given the above expressions for the non-zero connection components and the fact that the background de Sitter spacetime is maximally symmetric, one can determine the differential equation satisfied by the gravitational perturbation $h_{\mu \nu}$. To linear order in $h_{\mu \nu}$, one obtains the following wave equations in the de Sitter background
\bea\label{wave_eq_dS}
\bar{\square}\widetilde{h}_{\mu\nu}-\left[\bar{\nabla}_{\mu}B_{\nu}+\bar{\nabla}_{\nu}B_{\mu}-\bar{g}_{\mu \nu}\left(\bar{\nabla}_{\alpha}B^{\alpha}\right)\right]- \frac{2\Lambda}{3}\left(\widetilde{h}_{\mu \nu}-\widetilde{h}\bar{g}_{\mu\nu}\right)=-16\pi T_{\mu\nu}~,
\eea
where, as in the previous section, the ``bar'' denotes quantities evaluated for the background de Sitter spacetime. Also in the above expression, we have used the trace-reversed perturbation $\widetilde{h}_{\mu\nu}:= h_{\mu\nu}-(1/2)\bar{g}_{\mu\nu}h$ and its covariant divergence $B_{\mu}:=\bar{\nabla}_{\alpha}\widetilde{h} ^{\alpha}_{\mu}$. 

The above equation for the gravitational perturbation looks complicated and thus we need to impose an appropriate gauge condition in order to simplify it further. In the case of asymptotically flat background, one often chooses the Lorenz gauge condition, $B_{\mu}=0$, to simplify the perturbation equations. However, in the present context it is possible to choose another gauge, which will simplify the above wave equation considerably. 

For that purpose, let us concentrate on the wave equation associated with the spatial components $\widetilde{h}_{ij}$ of the gravitational perturbation. We choose the following wave gauge condition
\begin{equation} \label{wave gauge}
B_{\mu}:=f(t)\widetilde{h}_{0\mu}~,
\end{equation} 
where $f(t)$ is an arbitrary function. For this choice of $B_{\mu}$, taking a cue from \ref{wave_eq_dS}, the linearized wave equation for $\widetilde{h}_{ij}$ becomes
\begin{align} \nn \label{15VII20.1}
-\partial_{0}^{2} \widetilde{h}_{ij}&+e^{-2Ht} \left(\delta^{mk} \partial_{m} \partial_{k} \widetilde{h}_{ij}\right)+ H \partial_{0} \widetilde{h}_{ij}+ 2 H^{2} \widetilde{h}_{ij}-\delta_{ij}\Big[-2H^{2}e^{2Ht} \bar{g}^{kl} \widetilde{h}_{kl} 
\nonumber
\\
&+f(t)He^{2Ht}\widetilde{h}_{00}+e^{2Ht}(df/dt)\widetilde{h}_{00}+ e^{2Ht} f(t) \partial_{0}\widetilde{h}_{00}- f(t)\partial^{k} \widetilde {h}_{0k}\Big]
\nonumber
\\
&-\left(f(t)+2H\right)\left(\partial_{i} \widetilde{h}_{0j}+\partial_{j} \widetilde{h}_{0i}\right)=-16\pi T_{ij}~.
\end{align}
We observe that $f(t)=-2H$ will simplify it considerably. Writing down this gauge condition explicitly in terms of the trace reversed gravitational perturbation, we obtain $\bar{\nabla}_{\alpha}\widetilde{h}^{\alpha}_{\mu}=-2H \widetilde{h}_{0\mu}$. Imposing this condition, finally the wave equation for $\widetilde{h}_{ij}$ reads as
\bea 
-\partial_{0}^{2}\widetilde{h}_{ij}+H\partial_{0} \widetilde{h}_{ij}+e^{-2Ht}\left(\delta^{mk} \partial_{m}\partial_{k} \widetilde{h}_{ij}\right)+2H^{2}\widetilde{h}_{ij}=-16 \pi T_{ij}~.
\eea
Note that the wave equation for $\widetilde{h}_{ij}$ decouples from the other components of the gravitational perturbation. 

Further, expanding out $(\bar{\square}\widetilde{h}_{00})$ for the background de Sitter spacetime in the cosmological coordinates and using the gauge condition $B_{\mu}=-2H\widetilde{h}_{0\mu}$, introduced above, the evolution equation for the time-time component of the gravitational perturbation $\widetilde{h}_{00}$ becomes
\bea \label{15VII20.2}
-\partial_{0}^{2} \widetilde{h}_{00} + e^{-2Ht} \delta^{ij} (\partial_{i} 
\partial_{j} \widetilde{h}_{00}) -3H\partial_{0} \widetilde{h}_{00}-2H^{2} \widetilde{h}
_{00} - 2H^{2} e^{-2Ht} \delta^{ij} \widetilde{h}_{ij}=-16\pi T_{00}~.
\eea
Unfortunately, unlike the wave equation for the spatial part of the gravitational perturbation, the above wave equation for the $\widetilde{h}_{00}$ is a coupled differential equation. However, it is possible to decouple the spatial and the temporal part of the gravitational perturbation by introducing a redefined perturbation variable in favour of $\widetilde{h}_{00}$. The first step is to take the trace of \ref{15VII20.1} with respect to the flat spatial metric, which yields
\bea \label{15VII20.3}
-\partial_{0}^{2} \big({e^{-2Ht}\delta^{ij} \widetilde{h}_{ij}}\big)-3H\partial_{0}\big({e^{-2Ht}\delta^{ij} \widetilde{h}_{ij}}\big) +e^{-2Ht}\partial_{k}\partial^{k} \big({e^{-2Ht}\delta^{ij} \widetilde{h}_{ij}}\big)=-16 \pi  e^{-2Ht} {\delta^{ij}T_{ij}}~.
\eea
As a next step, we define a new perturbation variable 
\begin{equation}
    \widetilde{\mathcal{H}}:= \widetilde{h}_{00}+ {e^{-2Ht}} (\delta^{ij}\widetilde{h}_{ij})~.
\end{equation} 
Subsequently, summing up \ref{15VII20.2} and \ref{15VII20.3}, and using the definition for the gravitational perturbation $\widetilde{\mathcal{H}}$ given above, we obtain the following wave equation for $\widetilde{\mathcal{H}}$
\bea \label{30XI20.2}
-\partial_{0}^{2} \widetilde{\mathcal{H}}+ e^{-2Ht} \delta^{ij} \partial_{i} \partial_{j} \widetilde{\mathcal{H}} -3 H \partial_{0} \widetilde{\mathcal{H}} -2H^{2} \widetilde{\mathcal{H}}=-16\pi \left(T_{00}+{e^{-2Ht}}\delta^{ij} T_{ij}\right)~.
\eea
It is clear that the above wave equation for $\widetilde{\mathcal{H}}$ is decoupled, \emph{i.e.}, it depends on $\widetilde{\mathcal{H}}$ alone and not on other perturbation variables, as desired. Finally, the evolution equation for the temporal-spatial part of the perturbation, \emph{i.e.}, for $\widetilde{h}_{0i}$, takes the following form
\bea \label{26XI20.1}
-\partial_{0}^{2} \widetilde{h}_{0i} + e^{-2Ht}\delta^{jk} \partial_{j}\partial_{k} \widetilde{h}_{0i} - H \partial_{0}\widetilde{h}_{0i}=-16\pi T_{0i}.
\eea
Therefore, we have decoupled all the components of the gravitational perturbation, \emph{i.e.}, purely spatial part $\widetilde{h}_{ij}$, spatial-temporal part $\widetilde{h}_{0i}$ and $\widetilde{\mathcal{H}}$, a combination of purely temporal part and spatial part. This provides the desired wave equations for the gravitational perturbations around the de Sitter spacetime in cosmological coordinates. 

Concluding, as an aside remark, the wave gauge condition $\bar{{\nabla}}_{\alpha} \widetilde{h}^{\alpha}_{\mu}=-2H \widetilde{h}_{0\mu}$ in the cosmological coordinates becomes
\bea \label{wavegauge1}
-\partial_{0}\widetilde{h}_{0\mu}+e^{-2Ht}\partial^{j} \widetilde{h}_{j\mu}-H\delta^{0}_{\mu} \bar{g}^{kl} \widetilde{h}_{kl}- H \widetilde{h}_{0\mu}=0~.
\label{9VII20.1}
\eea
As it will turn out, it will play an important role in the subsequent section, where we would like to determine the corresponding residual gauge and highlight its connection with the multipole symmetry vector fields.

\subsection{Residual gauge transformations and multipole symmetries} \label{LinearRes}

In section \ref{Sec.4}, we have derived the multipole symmetry vector field $\xi^{\mu}$, see \ref{killing_FRW2}, in the cosmological coordinates, which respect the harmonic gauge condition of the de Sitter spacetime and preserves the relevant asymptotic fall-off condition, as in \ref{25XII20.1}. It is instructive to see any connection between the multipole symmetries and the residual diffeomorphism vector fields preserving the wave gauge condition \ref{wave gauge}-\ref{wavegauge1}, used in deriving the wave equations for linear gravitational perturbation in the cosmological coordinates. 

Suppose the wave gauge condition $\bar{\nabla}_{\alpha} \widetilde{h}^{\alpha}_{\mu}=-2H \widetilde{h}_{0\mu}$ is preserved by the diffeomorphism generating vector field $\xi^{\mu}$,\footnote{As of now, this vector field is completely different from the multipole symmetry vector in \ref{killing_FRW2}. Though, for convenience, we are using the same symbol to denote the diffeomorphism vector field in the present context as well.} whose form we would like to determine. First of all, note that under the transformation, $x^{\mu}\to x^{\mu}+\xi^{\mu}$, the trace-reversed gravitational perturbation transforms as
\bea \label{30XI20.1}
\delta_{\xi}\widetilde{h}_{\mu\nu}=\bar{\nabla}_{\mu}\xi_{\nu}+\bar{\nabla}_{\nu} \xi_{\mu}-\bar{g}_{\mu\nu}\left(\bar{\nabla}_{\alpha} \xi^{\alpha}\right)~.
\eea
The covariant derivatives appearing in the above expression are in the background de Sitter spacetime; expanding these derivatives in the cosmological coordinate yields
\bea 
\delta_{\xi}\widetilde{h}_{\mu\nu}=\partial_{\mu}\xi_{\nu}+\partial_{\nu}\xi_{\mu}-2H\left(\delta^{0}_{\nu}\xi_{\mu}+ \delta^{0}_{\mu} \xi_{\nu}\right)+H \xi_{0}\bar{g}_{\mu\nu}+ 2H \xi_{0} \delta^{0}_{\mu} \delta^{0}_{\nu}-\bar{g}_{\mu\nu} \bar{g}^{\alpha \beta} \partial_{\alpha} \xi_{\beta}~.
\eea
Thus, the wave gauge condition ${\bar \nabla}_{\alpha} \widetilde{h}^{\alpha}_{\mu}=-2H\widetilde{h}_{0\mu}$, will also be modified under the above diffeomorphism. As we want to preserve it, the vector field $\xi^{\mu}$ must satisfy the following differential equation
\bea
\bar{\square}\xi_{\beta}+ 4 H\partial_{0}\xi_{\beta}+2H^{2} \xi_{\beta} -2H^{2} \xi_{0}\delta^{0}_{\beta} +2H \delta^{0}_{\beta}\bar{g}^{\mu\nu}\partial_{\mu}\xi_{\nu}-\bigg(\frac{2H}{a^{2}}\bigg) \delta^{0}_{\beta} \delta^{ij} \partial_{i}\xi_{j}=0~.
\eea
Using explicitly the metric for the de Sitter spacetime in the cosmological coordinates, the above differential equation can be reduced to a partial differential equation 
\bea \label{30XI20.3}
\bar{g}^{\mu\nu}\partial_{\mu}\partial_{\nu} \xi_{\beta}+H\partial_{0}\xi_{\beta}+2 H^{2} \xi_{\beta}-2H^{2}\delta^{0}_{\beta} \xi_{0}-2H\delta^{0}_{\beta}\partial_{0}\xi_{0}=0~.
\eea
In addition to the above condition on $\xi^{\mu}$, we would like to see if this vector field can also be used to eliminate the $\widetilde{h}_{0i}$ and $\widetilde{\mathcal{H}}$ components of the gravitational perturbation. The $\widetilde{h}_{0i}$ component of the gravitational perturbation transforms under diffeomorphism as
\bea\label{h0i_change}
\delta_{\xi}\widetilde{h}_{0i}=\partial_{0}\xi_{i}+\partial_{i}\xi_{0}-2H\xi_{i}~.
\eea
In order to proceed further, we have to take various derivatives of the residual wave gauge condition. Firstly, taking the time derivative of the spatial component of the residual gauge condition in \ref{30XI20.3}, we obtain
\bea \label{30XI20.4}
\bar{g}^{\mu\nu} \partial_{\mu}\partial_{\nu} \partial_{0} \xi_{i}=2H\bar{g}^{kl}\partial_{k}\partial_{l} \xi_{i}-H\partial_{0}^{2}\xi_{i}-2H^{2} \partial_{0}\xi_{i}~.
\eea
Similarly, taking the spatial derivative of the temporal component of the residual gauge condition in \ref{30XI20.3}, we obtain
\bea \label{30XI20.5}
\bar{g}^{\mu\nu} \partial_{\mu}\partial_{\nu} \partial_{i}\xi_{0}=H\partial_{0}\partial_{i}\xi_{0}~.
\eea
Hence, we obtain the following result for the change in the spatial-temporal component of the gravitational perturbation under diffeomorphism
\begin{align} \label{30XI20.6}
\bar{g}^{\mu\nu}\partial_{\mu}\partial_{\nu}\delta_{\xi}\widetilde{h}_{0i}&= \bar{g}^{\mu\nu}\partial_{\mu}\partial_{\nu} (\partial_{0}\xi_{i}+\partial_{i}\xi_{0}-2H\xi_{i})
\nonumber
\\
&=2H\bar{g}^{kl}\partial_{k}\partial_{l} \xi_{i}-H\partial_{0}^{2}\xi_{i}-2H^{2} \partial_{0}\xi_{i}+H\partial_{0}\partial_{i}\xi_{0}+2H\left(H\partial_{0}\xi_{i}+2 H^{2} \xi_{i}\right)
\nonumber
\\
&=H\partial_{0}^{2}\xi_{i}+H\partial_{0}\partial_{i}\xi_{0}-2H^{2}\partial_{0}\xi^{i}~,
\end{align}
where in arriving at the first line we have used the identities derived in \ref{30XI20.4} and \ref{30XI20.5}, respectively. Further, in the second line we have used the result $\bar{g}^{kl}\partial_{k}\partial_{l} \xi_{i}=\partial_{0}^{2}\xi_{i}-H\partial_{0}\xi_{i}-2H^{2}\xi_{i}$, which follows from \ref{30XI20.3}. Thus, using \ref{h0i_change}, we finally obtain the differential equation for the change in the spatial-temporal component of the gravitational perturbation under diffeomorphism 
\bea
\bar{g}^{\mu\nu}\partial_{\mu}\partial_{\nu}\left(\delta_{\xi}\widetilde{h}_{0i}\right)-H\partial_{0} \left(\delta_{\xi}\widetilde{h}_{0i}\right)=0~,
\eea
which is the same of the wave equation satisfied by $\widetilde{h}_{0i}$ in the absence of any source; see \ref{26XI20.1}. Hence using the diffeomorphism vector field $\xi^{\alpha}$, which preserves the wave gauge condition, we can set $\widetilde{h}_{0i}=0$, outside the matter source. 

Again, under diffeomorphism, the combination $\widetilde{\mathcal{H}}$ of purely temporal and purely spatial part of the gravitational perturbation, transforms as
\bea \nn
\delta_{\xi}\widetilde{\mathcal{H}}&=&\delta \widetilde{h}_{00}+\bar{g}^{ij}\delta \widetilde{h}_{ij}=4\partial_{0}\xi_{0}~,
\eea
where we have used \ref{30XI20.1} to compute ${\delta \widetilde h_{00}}$ and $\bar{g}^{ij}\delta{\widetilde h_{ij}}$. Further, using the differential equation satisfied by $\xi^{\alpha}$, preserving the wave gauge condition, it can also be shown that $\delta_{\xi}\widetilde{\mathcal{H}}$ satisfies the following differential equation
\bea
\bar{g}^{\mu\nu}\partial_{\mu}\partial_{\nu}\left(\delta_{\xi}\widetilde{\mathcal{H}}\right)
-3H\partial_{0}\left(\delta_{\xi} \widetilde{\mathcal{H}}\right)-2H^{2}\delta_{\xi}\widetilde{\mathcal{H}}=0~,
\eea
which, upon comparison with \ref{30XI20.2}, turns out to be identical to the wave equation satisfied by the perturbation variable $\widetilde{\mathcal{H}}$ outside the source. Thus, we can also set $\widetilde{\mathcal{H}}=0$ outside the source using the diffeomorphism $\xi^{\mu}$, while preserving the gauge condition. 

To summarize, in addition to \ref{30XI20.3}, we also demand the conditions $\widetilde{h}_{0i}=0=\widetilde{\mathcal{H}}$ should be preserved under diffeomorphism  $\xi^{\mu}$, namely $\delta_{\xi} \widetilde{\mathcal{H}}=0$ and $\delta_{\xi} \widetilde{h}_{0i}=0$, to obtain
\begin{subequations}
\begin{align}
\partial_{0}\xi_{0}&=0~,
\\
\partial_{0} \xi_{i}+\partial_{i} \xi_{0}-2H \xi_{i}&=0~.
\end{align}
\end{subequations}
These differential equations can be immediately solved, yielding 
\begin{subequations}
\begin{align}
\xi_{0}&=-\epsilon(\bm{x})~,
\\ 
\xi_{i}&=e^{2Ht}\Big[\zeta_{i}(\bm{x})+\frac{1}{2H}\left(1-e^{-2Ht}\right)\partial_{i}\epsilon(\bm{x})\Big]~.
\end{align}
\end{subequations}
Raising the indices of both the components of the vector field $\xi^{\mu}$, we immediately obtain 
\bea
\xi^{0}=\epsilon(\bm{x})~,\qquad
\xi^{i}=\delta^{ij} \zeta_{j}(\bm{x})+\frac{1}{2H}\left(1-e^{-2Ht}\right)\delta^{ij}\partial_{i}\epsilon(\bm{x})~.
\eea
As one can explicitly verify, this is identical to the multipole symmetry vector derived earlier in the context of harmonic gauge; see \ref{killing_FRW2}. This demonstrates the internal consistency of our analysis and the relevance of the multipole symmetry vector field. They do not only obey the asymptotic fall-off conditions and preserve the harmonic gauge condition, but they further can be used to eliminate the time-space component $\widetilde{h}_{0i}$ and $\widetilde{\mathcal{H}}$, a suitable combination of the time-time and space-space component of the gravitational perturbation around the de Sitter background in cosmological coordinates.

\section{Multipole structure of the Kerr-de Sitter black hole}\label{Sec:KdS}

The formalism for computing the multipole moments of a compact object in asymptotically de Sitter spacetime follows closely the method in \cite{Compere:2017wrj} for asymptotically flat spacetimes. 
In this section, we will compute the multipole moments of the Kerr-de Sitter (KdS) black hole spacetime, which will allow us to show explicitly how the mass and spin multipole moments for the KdS black hole differ from the Geroch-Hansen moments for Kerr black hole. We believe that the explicit expressions for the mass and spin multipole moments of the KdS black hole are new in the literature.

The key object to compute the multipole moments is the Barnich-Brandt charge \cite{Barnich:2001jy}, or equivalently the Abbott-Deser charge \cite{Abbott:1981ff}. In four-dimensional General Relativity, the infinitesimal surface charge $ \delta \mathcal{Q}_{\xi}[h;\bar{g}]$ associated to the vector field $\xi$ and the linearized solution $h_{\mu\nu}$ around a background spacetime $\bar{g}_{\mu\nu}$ is 
\begin{equation}\label{BBcharge}
\delta \mathcal{Q}_{\xi}[h;\bar{g}]:=\frac{1}{8\pi }\int_S \bm{k}_{\xi}[h; \bar{g}]=\frac{1}{32\pi }\int_S \sqrt{-\bar{g}}~ k^{\mu\nu}_{\xi}[h; \bar{g}]~ \epsilon_{\mu\nu\alpha \beta}~dx^{\alpha}\wedge dx^{\beta}~,
\end{equation}
where, the surface charge density $k^{\mu\nu}_{\xi}$ is given by \cite{Barnich:2001jy}
\begin{equation}\label{chargedensity}
k^{\mu\nu}_{\xi}[h; \bar{g}]:=\xi^{\nu}\left(\bar{\nabla}^{\mu}h-\bar{\nabla}_{\sigma}h^{\mu\sigma} \right) + \xi_{\sigma}\bar{\nabla}^{\nu}h^{\mu\sigma} + \frac{1}{2} h \bar{\nabla}^{\nu}\xi^{\mu} - h^{\rho\nu}\bar{\nabla}_{\rho}\xi^{\mu} + \frac{1}{2} h^{\sigma \nu}\left(\bar{\nabla}^{\mu}\xi_{\sigma} + \bar{\nabla}_{\sigma}\xi^{\mu} \right)~.
\end{equation}
In the case of the KdS black hole, the background metric $\bar{g}_{\mu\nu}$ is the metric for the de Sitter spacetime, while $h_{\mu\nu}$ is the linearized solution obtained by varying the parameters of the KdS black hole. We will first provide a brief review of the KdS black hole spacetime, since it will be extremely useful for our subsequent computation of the multipole moments from the charges associated with multipole symmetry vector fields. 

\subsection{Kerr-de Sitter black hole}

KdS black hole is an exact solution to the Einstein's equations with a positive cosmological constant $\Lambda=3H^{2}>0$ \cite{Carter:1968ks,GIBBONS200549}. The metric of the KdS black hole spacetime reads as\footnote{See, \emph{e.g.} the $\Lambda>0$ counterpart of the metric in \cite{Olea:2005gb} and references therein for a complete account of the Kerr-dS and Kerr-AdS black hole thermodynamics. Sometimes, the KdS metric is written with the time rescaled by $\Xi$; see, \emph{e.g.}, \cite{Akcay:2010vt}.}
\begin{equation}\label{metric:KdS}
ds^2 = -\frac{\Delta_r}{\rho^2}\left(dt - \frac{a}{\Xi}\sin^2 \theta d\phi \right)^2 + \frac{\rho^2}{\Delta_r}dr^2+ \frac{\rho^2}{\Delta_{\theta}}d\theta^2 +\frac{\Delta_{\theta}}{\rho^2}\sin^2 \theta\left(adt -\frac{r^2+a^2}{\Xi}d\phi \right)^2,
\end{equation}
where the functions in the metric components are given by
\begin{subequations}
\begin{align}
\Delta_r &= \left(r^2+a^2\right)\left(1-H^2 r^2 \right)-2Mr,\\
\Delta_{\theta} &= 1+H^2 a^2 \cos^2 \theta,\\
\rho^2 &= r^2 + a^2 \cos^2 \theta,\\
\Xi &= 1+a^2H^2 .
\end{align}
\end{subequations}
The KdS black hole is described by three parameters: the mass $M$, the spin $a$, and the Hubble constant $H$. While we allow the mass $M$ and the spin $a$ to vary, we keep the Hubble constant $H$ fixed. The de Sitter spacetime (\emph{i.e.}, the background spacetime) is recovered for $M=0=a$, while the Kerr black hole is recovered for $H=0$. 

The KdS black hole spacetime is considered as a two-parameter family, since as the parameters $M$ and $a$ change, the metric changes its configuration. The linearized perturbation $h_{\mu\nu}$, necessary for the computation of the charges can be considered as tangent to the space of the metric configurations and computed as $\delta g_{\mu\nu}:= h_{\mu \nu} = (\partial_M g_{\mu\nu})\delta M + (\partial_a g_{\mu\nu})\delta a$. This sets the stage for our subsequent application of our prescription to the KdS black hole and hence to compute its mass and spin multipole moments.

\subsection{Multipole moments of the Kerr-de Sitter black hole}

In what follows, we will show how to compute in practice the multipole moments of the KdS black hole. Here are the main steps:
\begin{itemize}
    \item First of all, one starts by decomposing in spherical harmonics the multipole symmetry vectors in \ref{MultipoleSymm}. The vector fields $\xi=\{\bm{K}_{\epsilon},\bm{L}_{\epsilon}\}$ are named, respectively, the mass and spin multipole symmetries because they generate the mass and spin multipole moments of the compact object in the asymptotically de Sitter spacetimes\footnote{We also assume that, in the asymptotic regime, the coordinates of the KdS black hole coincide with that of the cosmological coordinates.
    };
    \item Then, we will choose $h_{\mu\nu}$ to be the linearized perturbation of the KdS metric with $\bar{g}_{\mu\nu}$ being the dS background and shall compute the surface charge density, using \ref{chargedensity}. Integration of the same over a generic 2-sphere, yields the surface charges, according to \ref{BBcharge}. One thus gets the $(l,m)$ modes of the infinitesimal charge $\delta \mathcal{Q}^{lm}_{\xi}$;
    \item The infinitesimal charge mode $\delta \mathcal{Q}^{lm}_{\xi}$, derived above, is, in general, a function of $t$ and $r$. One considers a $t=\textrm{constant}$ time slice and computes the large radius expansion of the mode $\delta \mathcal{Q}^{lm}_{\xi}$. Only its finite part (FP) is retained, namely only the coefficient of $\mathcal{O}(r^0)$ is kept;
    \item In the final step, one integrates the $\text{FP}~ \delta \mathcal{Q}^{lm}_{\xi}$ over the solution parameters, \emph{i.e.}, over the parameters $m$ and $a$, to get the multipole moments $M_{\xi}^{lm}$ of the solution. In formul\ae,
    \begin{equation}\label{mult_mom}
    M^{lm}_{\xi} = \underset{\begin{subarray}{c}
    r\to\infty \\
    t=\text{constant}
    \end{subarray}}{\text{FP}} \mathcal{Q}^{lm}_{\xi}.
    \end{equation}
\end{itemize}
\subsubsection{Spin multipole moments}

The first step in the computation of the spin multipole moments is to recall the symmetry vector associated with spin multipole symmetry, $\bm{L}_{\epsilon} = -{\bm r} \times {\bm \nabla} \epsilon({\bm x})$, written in the same coordinates as that of the KdS metric. We call $\bm{L}_{\epsilon}$ the spin multipole symmetry since it will generate the spin multipole moments. The spherical harmonic decomposition of the spin multipole symmetry vector field is given by\footnote{We choose the convention that $\bm{L}_{10} = - \partial_{\phi}$.} 
\begin{equation}\label{spinL}
\bm{L}_{lm} = \mathcal{L}_l~ r^{l-1}\left(~^{\rm B}Y^{\theta}_{lm}\partial_{\theta}  + \frac{1}{\sin \theta} ~^{\rm B}Y^{\phi}_{lm}\partial_{\phi}\right)~,
\end{equation}
where the magnetic-type harmonic vector field $~^{\rm B}\bm{Y}_{lm}$ is defined by
\begin{equation}
~^{\rm B}\bm{Y}_{lm} = \frac{1}{\sqrt{l(l+1)}} ~\bm{r} \times \bm{\nabla} Y_{lm}, \quad Y_{lm} = (-1)^m \sqrt{\frac{2l+1}{4\pi} \frac{(l-m)!}{(l+m)!}} e^{im\phi}P_{lm}(\theta)~,
\end{equation}
with $P_{lm}(\theta)$ being the associated Legendre polynomials and $Y_{lm}$ the spherical harmonics. The normalization factor, $\mathcal{L}_{l}$, at this stage of the discussion, is not fixed. However, we decide to adjust it in such a way that we recover the spin multipole moments of the Kerr black hole in the limit $H \to 0$. Such requirement implies that the normalization factor takes the following form
\begin{equation}
\mathcal{L}_{l}=\frac{8\sqrt{\pi}}{3}\sqrt{\frac{l(2l+1)}{l+1}}\frac{(2l-1)!!}{(l+1)!}~.
\end{equation}
We now use the expression of the surface charge from \ref{BBcharge} and the explicit expression of the spin multipole symmetry from \ref{spinL} to compute the spin multipole moments $\mathcal{S}_{l}$ of the KdS black hole, using \ref{mult_mom}. It is important to recall that the charge mode $\mathcal{Q}^{lm}_{L}$ is non-vanishing for $l$ odd and $m=0$, and so is the spin multipole moment $\mathcal{S}_{l}$. 

The angular momentum, \emph{i.e.} the spin dipole moment $l=1$, reads as
\begin{equation}
\mathcal{S}_{1} = \frac{Ma}{\left(1+a^2H^2\right)^2}~,
\end{equation}
in agreement with the known results in the literature \cite{ABKI, Chrusciel:2015sna, Olea:2005gb,Akcay:2010vt}. Moreover, in the limit $H\to 0$, we recover the angular momentum of the Kerr black hole. 

The higher spin multipole moments are computationally more involved. We report the exact expressions of the first spin multipole moments in Appendix \ref{Appspin}. It is more instructive to compute their expressions for small values of the Hubble constant:
\begin{subequations}
\begin{align}
\mathcal{S}_{3} &= -M a^3  \left[1 -\frac{28}{15} a^2 H^2 + \frac{25}{9} a^4 H^4 + \mathcal{O}(H^6)\right]~,\\
\mathcal{S}_{5} &= +M a^5  \left[1 - \frac{118}{63} a^2 H^2 + \frac{2891}{1053} a^4 H^4 + \mathcal{O}(H^6)\right]~,\\
\mathcal{S}_{7} &= -M a^7  \left[1 - \frac{17}{9} a^2 H^2 + \frac{513}{187} a^4 H^4+ \mathcal{O}(H^6)\right]~,\\
\mathcal{S}_{9} &= +M a^9  \left[1 - \frac{314}{165} a^2 H^2 + \frac{3751}{1365} a^4 H^4+ \mathcal{O}(H^6)\right]~.
\end{align}
\end{subequations}
For $H \to 0$, they reproduce the well-known Geroch-Hansen's formul\ae~ for the spin multipole moments of the Kerr black hole, namely,
\begin{equation}
\underset{H \to 0}{\lim} ~\mathcal{S}_{2l+1} = (-1)^l M a^{2l+1}~.
\end{equation} 
Thus the spin multipole moments derived here satisfies the result that all even spin moments are identically zero. This is because of the reflection symmetry of the KdS spacetime about the equatorial plane. Note that all the corrections over and above the multipole moments of the Kerr black hole are dependent on the dimensionless combination $(aH)$, which will be small if we consider $H^{-1}$ as the age of our universe. Thus, the effect of the cosmological constant on the spin multipole moments are negligible. We will analyze the situation for the mass multipole moments in a while. 

\subsubsection{Mass multipole moments}

The mass multipole symmetry vector reads as $\bm{K}_{\epsilon} = \epsilon(\bm{x}) \partial_{t}+(1/2H)(1-e^{-2Ht})\bm{\nabla}\epsilon(\bm{x})-Hx^{i}\partial_{i}$, which admits a spherical harmonic decomposition of the form
\begin{equation} \label{Kharmonic}
\bm{K}_{lm}=\mathcal{K}_{l}~r^l\left(Y_{lm}\partial_t + \frac{1}{r}\chi^{r}_{lm}\partial_r + \frac{1}{r^2}\chi^{\theta}_{lm}\partial_{\theta} + \frac{1}{r^2 \sin \theta}\chi^{\phi}_{lm}\partial_\phi \right) 
-H\delta_{l1}\partial_r~.
\end{equation}
The spatial vector field $\bm{\chi}_{lm}$, whose components appear explicitly in the above expression, is a certain linear combination of the electric-type vector harmonic, defined as $~^{\rm E}\bm{Y}_{lm} = r {\bm \nabla} Y_{lm} $, and the radial-type vector harmonic, $~^{\rm R}\bm{Y}_{lm} = \bm{n}Y_{lm}$, such that
\begin{equation}
\bm{\chi}_{lm}=\frac{1}{2H}\left(1-e^{-2Ht}\right)\left(\sqrt{l(l+1)}~^{\rm E}\bm{Y}_{lm}+l ~^{\rm R}\bm{Y}_{lm} \right)~.
\end{equation}
After computing the surface charge given in \ref{BBcharge}, associated to the mass multipole symmetry, the mass multipole moments are non-vanishing for $l$ even and $m=0$. Thus the mass multipole moments are denoted by $\mathcal{M}_{l}$. Moreover, we demand that the normalization factor $\mathcal{K}_l$ in the mass multipole symmetry is such that we recover the Geroch-Hansen's mass moments for Kerr black hole. This implies that
\begin{equation}\label{Kl}
\mathcal{K}_l=\sqrt{\pi}\sqrt{2l+1}\left(\frac{2^{l/2+1}}{N_{l/2+1}}\right)\frac{(2l-1)!!}{l!}~,
\end{equation}
and $N_l$ obeys the following recursive relation for $l\geq 2$,
\begin{equation}
N_{l+2} = -2\frac{(17-12l)N_{l+1}+10(l-2)N_l}{7l-9}~,
\end{equation}
with initial conditions $N_1=1$ and $N_2=4$.

Following this strategy outlined above, the mass of the KdS black hole, \emph{i.e.}, the monopole $l=0$ mode takes the form
\begin{equation}
\mathcal{M}_0 = \frac{M}{1+a^2H^2}~,
\end{equation}
which reproduces the de Sitter analogue of the mass for the Kerr-AdS black hole obtained in \cite{Olea:2005gb},\footnote{Notice that we could have also rescaled the time coordinate by $\Xi$ in the KdS metric, see \ref{metric:KdS}, and obtain $\mathcal{M}_0 = M/(1+a^2H^2)^2$ to match with some earlier results in the literature \cite{Chrusciel:2015sna, ABKI, Akcay:2010vt}. This is a trivial modification in the computation and it does not affect the validity of our prescription. Moreover, though the $H \to 0$ limit is left untouched, any time-dependent rescaling affects $\mathcal{O}(H^n)$ corrections to the Geroch-Hansen's mass moments.} and gives the correct expression for the mass of the Kerr black holes for $H=0$.
The mass quadrupole, as expected, is time independent and it is given by
\begin{equation}
\mathcal{M}_2 = -\frac{Ma^2}{1+a^2H^2}~.
\end{equation}
The exact expressions of the higher mass multipole moments can be found in Appendix \ref{Appmass}. For small values of the Hubble constant, they read as 
\begin{subequations}
\begin{align}
\mathcal{M}_4 &= +Ma^4\left[ 1 + \frac{4}{5}Ht - \frac{47a^2 + 44t^2}{55} H^2 + \mathcal{O}(H^3)\right]~,\\
\mathcal{M}_6 &= -Ma^6 \left[1+\frac{6}{7} Ht -\frac{7a^2+30t^2}{35}H^2 + \mathcal{O}(H^3) \right]~,\\
\mathcal{M}_8 &= +Ma^8 \left[ 1+\frac{8}{9} Ht +\frac{9a^2-152t^2}{171}H^2 + \mathcal{O}(H^3) \right]~.
\end{align}
\end{subequations}
As evident from the above expressions, for $H \to 0$, we recover the well-known Geroch-Hansen's formul\ae~ for the mass multipole moments of the Kerr black hole
\begin{equation}
\underset{H \to 0}{\lim} ~\mathcal{M}_{2l} = (-1)^{l}Ma^{2l}.
\end{equation}
Notice that, owing to the reflection symmetry of the KdS spacetime about the equatorial plane, only the even order mass moments are non-zero. 

We conclude this section with an interesting outcome from the mass multipole moments computation. While the $H\to 0$ limit, or the static limit, of $\mathcal{M}_{2l}$ gives the Hansen's mass moments for the Kerr black hole, there exists another limit, which may be of interest. It is indeed true that except for the first two, all the other higher order mass moments are time dependent and the time variable enters the expressions via the exponential function $e^{-2Ht}$, typical of the de Sitter dynamics; see Appendix \ref{Appmass}. One can therefore perform the late-time limit, \emph{i.e.}, $Ht \to \infty$. In this case, one has
\begin{subequations}
\begin{align}
   \underset{Ht \to \infty}{\lim} \mathcal{M}_2 &= -\frac{Ma^2}{1+a^2H^2}~,\\
   \underset{Ht \to \infty}{\lim} \mathcal{M}_4 &= +\frac{7}{5}\frac{Ma^4}{1+a^2H^2}~,\\
    \underset{Ht \to \infty}{\lim} \mathcal{M}_6 &= -\frac{10}{7}\frac{Ma^6}{1+a^2H^2}~,\\
   \underset{Ht \to \infty}{\lim} \mathcal{M}_8 &= +\frac{13}{9}\frac{Ma^8}{1+a^2H^2}~.
\end{align}
\end{subequations}
While the quadrupole mass moment is the same in both the static and late-time limits, higher mass multipole moments differ by a numerical factor. More precisely, they can be recast as
\begin{equation}
\underset{Ht \to \infty}{\lim} \mathcal{M}_{2l}=\frac{\tilde{\mathcal{K}}_{l}}{2l+1}\frac{(-1)^l Ma^{2l}}{1+a^2H^2},
\end{equation}
with $\tilde{\mathcal{K}}_{l}$ satisfying the recursive relation: $\tilde{\mathcal{K}}_{l+2} = -2\tilde{\mathcal{K}}_{l+1} - \tilde{\mathcal{K}}_{l}$, with initial conditions $\tilde{\mathcal{K}}_{1}=-3$, $\tilde{\mathcal{K}}_{2}=7$, $\tilde{\mathcal{K}}_{3}=-10$, for $l\geq 4$.
It is straightforward to define a different normalization factor in the mass multipole symmetry (see \ref{Kl}), such that $\mathcal{K}_l \to (2l+1)\mathcal{K}_l/\tilde{\mathcal{K}}_{l}$ to get rid of the numerical factor and obtain, in the late-time limit, the Hansen's mass moments rescaled by the factor $\Xi = 1+a^2H^2$. In the asymptotically flat case, for which $H \to 0$, the Hansen' mass moments are recovered.

\section{Discussion and concluding remarks}\label{conclusion}

We have addressed the problem of computing the gravitational multipole moments of a compact object, living in an asymptotically de Sitter spacetime. Since the standard approaches of computing the gravitational multipole moments relies heavily on the asymptotic flatness, they could not be used in computing the moments for asymptotically de Sitter spacetimes. We achieved our results by using the method proposed in \cite{Compere:2017wrj} to compute the multipole moments by means of Noether charges associated with specific residual harmonic gauge transformations. 

The application of the Noether charge technique to compute gravitational multipole moments in asymptotically de Sitter spacetime requires one to implement the following steps: (a) Expressing the de Sitter spacetime in harmonic gauge, (b) Finding out the symmetry vector field, generating residual gauge transformations in the de Sitter spacetime, expressed in harmonic gauge, (c) Checking the consistency of the symmetry vector field with the corresponding one associated with linear gravitational perturbations around the de Sitter background. All these lead to a unique vector field that depends on the spherical harmonic decomposition of a function $\epsilon(\bm{x})$, satisfying Laplace's equation. This vector field can be decomposed into three parts --- (i) $\bm{K}_{\epsilon}$, generating mass multipole moments, (ii) $\bm{L}_{\epsilon}$, generating spin multipole moments and (iii) $\bm{P}_{\epsilon}$, which does not provide any further independent multipole moments in General Relativity.    

Following this strategy, in section \ref{Harmonic} we provide the transformation of the de Sitter metric expressed in the cosmological coordinates to the harmonic coordinates. To our knowledge, such a transformation of the de Sitter metric to the harmonic coordinates has not been attempted before. Having transformed the de Sitter metric to the harmonic form, we derive the diffeomorphism vector field respecting the harmonic gauge as well as the asymptotic fall-off condition \ref{25XII20.1}. The expression for the resulting multipole symmetries are given in \ref{MultipoleSymm}. 
In addition, this vector fields can also be used to eliminate the non-dynamical components of the linear gravitational perturbation around the de Sitter background in cosmological coordinates. Incidentally, we provide an analysis of the linear gravitational perturbation of the de Sitter spacetime in the cosmological coordinates and the associated appropriate gauge condition, referred to as the wave gauge.

The above formalism allows us to compute the gravitational multipole moments of any compact object in asymptotically de Sitter spacetime. The  charges associated to the above-mentioned multipole symmetries are the key objects from which we can extract the multipole moments. As an example of this procedure, we consider the case of the Kerr-de Sitter black hole spacetime and following the prescription outlined in section \ref{Sec:KdS}, we compute the mass and spin multipole moments of the Kerr-de Sitter black hole. It turns out that the mass (monopole mass moment) and the angular momentum (dipole spin moment) reproduce earlier results in the literature. However, the higher order mass and spin multipole moments for Kerr-de Sitter spacetime do not exist in the literature, and are discussed in this work for the first time. It turns out that the spin moments involve additional corrections over and above the Kerr moments, depending on the dimensionless combination $aH$. Since for the present epoch $H\sim (\text{age~of~universe})^{-1}$, the corrections depending on various powers of the combination $aH$ are supposed to be negligible. On the other hand, except for the quadrupole mass moment, all the higher order mass moments are time dependent and depends on various powers of the combination $aH$ as well as $Ht$, over and above the Kerr mass moments. It is assuring that in the $H\to 0$ limit, we recover the mass and spin moments of the Kerr spacetime. The higher order mass moments, though complicated, in the late time limit ($Ht\to \infty$) takes a very simple form. Notice that the mass moments of the Kerr-de Sitter black hole in the late time limit is very much related to the Kerr mass moments, except for some overall normalization factor. It is worthwhile to mention that, even though $aH$ is small, since in the present epoch $Ht\sim \mathcal{O}(1)$, there can be significant departure of the mass moments from that of the Kerr black hole. Since these moments directly affect various gravitational wave observables, \emph{e.g.}, the energy emitted by gravitational waves, there can be some observational consequences, which we wish to explore in a future work.  

Let us finally conclude with some future directions. To date, the geometrical Geroch-Hansen formalism is applicable only to stationary and asymptotically flat spacetimes. A few attempts exist in the literature to extend the formalism to asymptotic non-flat cases, see \emph{e.g.} \cite{Mukherjee_2020}, where multipole moments of spacetimes with NUT charges has been computed. In line with these developments, it would be interesting to develop a formalism \`a la Geroch-Hansen for asymptotically (anti-)de Sitter spacetimes. Additionally, a method to compute the multipole moments for radiating spacetimes, following the approach of Thorne, is also non-existent in literature for asymptotically (anti-)de Sitter spacetimes. A possible extension of the same to the (anti-)de Sitter spacetimes will prove very useful for various gravitational-wave-related implications. These will also provide other independent methods, than the one presented in this work, to compute the multipole structure of the Kerr-(anti-)de Sitter black hole spacetime. It is also important to ask, whether the formalism developed here can be successfully applied to investigate the complex multipole structure, and its related properties, of neutron stars and exotic objects, such as boson stars and fuzzball configurations. Finally, it will be very interesting to investigate the relation between the multipole symmetries for asymptotically de Sitter spacetimes and the adiabatic modes in cosmology; see, \emph{e.g.}, \cite{Hinterbichler_2014,mirbabayi2016weinberg,Hamada_2018}. We will come back to these further applications elsewhere.

\section*{Acknowledgements}
The authors thank Geoffrey Comp\`ere, Ali Seraj and the anonymous referee for useful comments and feedback on the manuscript.
Research of S.C. is funded by the INSPIRE Faculty fellowship from the DST, Government of India (Reg. No. DST/INSPIRE/04 /2018/000893) and by the Start-Up Research Grant from SERB, DST, Government of India (Reg. No. SRG/2020/000409). The research of J.H. is supported in part by the Czech Science Foundation Grant 19-01850S. The research of R.O. is funded by the European Structural and Investment Funds (ESIF) and the Czech Ministry of Education, Youth and Sports (MSMT), Project CoGraDS - CZ.02.1.01/0.0/0.0/15003/0000437. 

\appendix

\section{de Sitter in harmonic coordinates}\label{App:harm}

In this Appendix, we provide the explicit coordinate transformations from static and global coordinates to harmonic ones for the de Sitter spacetime.

\subsection{From static to harmonic coordinates}

The de Sitter spacetime has already been expressed in the static coordinates $(T,R,\theta,\phi)$ in \ref{dS_static}:
\be
ds^{2}= -\big(1-H^{2}R^{2}\big) dT^{2} +\frac{dr^2}{\big(1-H^{2}R^{2}\big)}+R^2\left(d \theta^2+\sin^{2}\theta d\phi^{2}\right).
\ee
As one can explicitly check, the metric indeed solves the Einstein's field equations with a positive cosmological constant term, $\Lambda \equiv 3H^{2}$. It is also straightforward to check that the de Sitter metric in the static coordinates does not obey the harmonic gauge condition, \emph{i.e.}, $\partial_{\mu}\left( \sqrt{-g}g^{\mu\nu}\right)\neq 0$. In what follows, we will depict an explicit coordinate transformation taking the de Sitter spacetime from static coordinates to the harmonic ones.

The strategy to arrive at the harmonic coordinates is to first introduce the Cartesian coordinates, \emph{i.e.}, consider the following coordinate transformation 
\begin{equation}
    x = R \cos \phi\sin \theta, \quad y=R \sin\phi \sin \theta, \quad z=R \cos\theta, 
    \end{equation}
so that the line element becomes,
\begin{equation}\label{dS_Static_01}
ds^{2}=-\left(1-H^2R^2\right)dT^2+\left(\delta_{ij} + \frac{H^2R^2}{1-H^2R^2}n_i n_j\right)dx^i dx^j,
\end{equation}
where $n_{i}=x_{i}/R$ is the radial unit vector. The second and the final step consists of a coordinates transformation $x^\mu \to \bar{x}^{\mu}(x^{\alpha})$, such that $\Box_{g} \bar{x}^\mu =0$, where $g_{\mu \nu}$ is the metric given in \ref{dS_Static_01}. To achieve the second step, we introduce a new radial coordinate $\bar{r}=f(R)$, such that
\begin{align}
\bar{t}&=T~, \quad \bar{x}=f(R)\cos\phi \sin\theta~, \quad
\bar{y}=f(R) \sin\phi \sin\theta~,\quad \bar{z}= f(R) \cos \theta.
\end{align}
The harmonic gauge condition $\Box_{g}\bar{x}^{\mu}=0$, when applied on the above set of coordinates, yields the following second order ordinary differential equation for the function $f(r)$
\begin{equation}
\frac{d}{dR}\left[R^2\left( 1-H^2 R^2\right)\frac{df(R)}{dR}\right] - 2f(R)=0,
\end{equation}
whose most general solution is 
\begin{equation}
f(R) = \alpha \left(1+\frac{1}{H^2R^2}\right) + \frac{\beta}{4}\left[\left(1+\frac{1}{H^2R^2}\right) \tanh^{-1} \left( H R\right) -\frac{1}{HR}\right], \quad \alpha, \beta \in \mathbb{R}.
\end{equation}
In order fix the constants of integration $\alpha$ and $\beta$, we need to impose appropriate boundary conditions. We demand the following condition on the function $f(R)$, namely 
\bea
f(R=1/H) = 1/H, 
\eea 
which provides the following expressions for the coefficients: $\alpha= 1/(2H)$ and $\beta=0$. Thus, in the new system of coordinate $(\bar{t}, \bar{x}, \bar{y}, \bar{z})$, the de Sitter metric in the static patch reads,
\begin{equation}
ds^2 = -2\frac{\left( H\bar{r}-1\right)}{2H \bar{r}-1}d\bar{t}^2+ \frac{1}{2}\frac{d\bar{r}^2}{\left( H\bar{r}-1\right)\left(2 H\bar{r}-1\right)^2} + \frac{1}{H^2\bar{r}^2 \left( 2H\bar{r}-1\right)}\left( \delta_{ij} - \bar{n}_i\bar{n}_j\right)d\bar{x}^i d\bar{x}^j.
\end{equation}
As one can explicitly check that, through a lengthy but straightforward algebra, the above metric element indeed satisfies the harmonic gauge condition. 

\subsection{From global to harmonic coordinates}

Another useful coordinate system for de Sitter spacetime is the global coordinate system $(\tau,\chi,\theta,\phi)$ with the line element given by \ref{dS_global}:
\be
ds^{2}=-d\tau^2 +\frac{1}{H^{2}}\cosh^{2}(H\tau)\left[d\chi^2 +\mathrm{sin}^2\chi \left(d\theta^2 + \mathrm{sin}^2\theta d\phi^2\right)\right]. 
\ee
In global coordinates, we obtain $\sqrt{-g}=(1/H^{3})\cosh^{3}(H\tau)\sin^{2}\chi \sin\theta$ and hence one gets
\begin{align}
\sqrt{-g}g^{\mu \nu}=&\textrm{diag}\Big(-H^{-3}\cosh^{3}(H\tau)\sin^{2}\chi \sin\theta,H^{-1}\cosh(H\tau)\sin^{2}\chi \sin\theta,
\nonumber
\\
&\quad \qquad H^{-1}\cosh(H\tau)\sin \theta,H^{-1}\cosh(H\tau)\textrm{cosec}\theta \Big).
\end{align}
As one can check, the global coordinate system defined above does not satisfy the harmonic gauge condition, \emph{i.e.}, $\partial_{\mu}(\sqrt{-g}g^{\mu \nu})\neq 0$. Thus, we need to find out a new set of coordinates which satisfies the harmonic gauge condition and for that purpose, the following coordinate transformation becomes useful
\begin{align}
\bar{\tau}=f(\tau), \quad  \bar{x}=g(\chi)\sin \theta \cos \phi, \quad \bar{y}=g(\chi)\sin \theta \sin \phi, \quad \bar{z}=g(\chi)\cos\theta.
\end{align}
In order for these coordinates to obey the harmonic gauge condition, we need them to satisfy the following differential equation: $(1/\sqrt{-g})\partial_{\alpha}(\sqrt{-g}g^{\alpha \beta})\partial_{\beta}\bar{x}^{\mu}=0$, where $g_{\alpha \beta}$ is the metric introduced in \ref{dS_global}. From the time component of the above differential equation, we obtain the following ordinary differential equation for the function $f(\tau)$
\begin{align}
\frac{d^{2}f}{d\tau^{2}}+3H\tanh(H\tau)\frac{df}{d\tau}=0.
\end{align}
The first integral of the above differential equation yields, $(df/d\tau)=A~\textrm{sech}^{3}(H\tau)$, whose further integration yields
\begin{align}
\bar{\tau}=B+A\left[\frac{1}{H}\tan^{-1}\left\{\tanh\left(\frac{H\tau}{2}\right)\right\}+\frac{1}{2H}\textrm{sech}(H\tau)\tanh(H\tau)\right].
\end{align}
If we impose the condition $\bar{\tau}=0$, when $\tau=0$, then it follows that we can set the coefficients such that $A=1$ and $B=0$, thereby providing the desired relation between the harmonic gauge time coordinate $\bar{\tau}$ and the global time coordinate $\tau$. On the other hand, imposing the harmonic gauge condition on the coordinate $\bar{x}$, we obtain the following differential equation for the function $g(\chi)$ as
\begin{align}
\frac{d}{d\chi}\left(\sin^{2}\chi\frac{dg}{d\chi}\right)-2g=0.
\end{align}
with solution $g(\chi)=A\textrm{cosec}^{2}\chi+B \textrm{cosec}^{2}\chi \{(\chi/2)-(1/4)\sin(2\chi)\}$. Setting $A=1$ and $B=0$, we obtain the following line element for de Sitter spacetime in the global patch in harmonic coordinate system
\begin{align}
ds^{2}=-\cosh^{6}(H\tau)d\bar{\tau}^{2}+\frac{\cosh^{2}(H\tau)}{H^{2}\bar{r}^{3}}\left[\delta_{ij}+\left(\frac{4-3\bar{r}}{4(\bar{r}-1)}\right)\bar{n}_{i}\bar{n}_{j}\right]d\bar{x}^{i}d\bar{x}^{j},
\end{align}
where, $\bar{r}=\sqrt{\bar{x}^{2}+\bar{y}^{2}+\bar{z}^{2}}$ and $\bar{n}_{i}=(1/\bar{r})\bar{x}^{i}$. One can check that the metric element above indeed satisfies the harmonic gauge condition. Thus we have found out the set of coordinates in which even the global patch of the de Sitter spacetime can be converted to harmonic gauge.

\section{Mass and spin moments of Kerr-de Sitter black hole} \label{exactmoments}
We list the exact analytic expressions of the first mass and spin multipole moments of the Kerr-de Sitter black hole. The function $Li_2(z)$, appearing in the spin multipole moments, is the poly-logarithm function of order two. It can be represented by the power series $Li_2(z) = \sum_{k=1}^{\infty} z^k / k^2$.
The computation has been performed with the use of  \href{https://library.wolfram.com/infocenter/MathSource/4484/}{\emph{Riemanian Geometry and Tensor Calculus}} package developed by Sotirios Bonanos and the \href{https://ptm.ulb.be/gcompere/package.html}{\emph{Surface Charges}} package developed by Geoffrey Comp\`ere.

\subsection{Spin multipole moments} \label{Appspin}
Here are the exact expressions of the spin multipole moments $\mathcal{S}_{2l+1}$ for $l = \{1, 2, 3, 4\}$.
\begin{subequations}
\begin{align}
    \mathcal{S}_3 &= -Ma^3 \Bigg\{ \frac{525 +1038a^2H^2+601a^4H^4}{96 a^4 H^4 \left(1+a^2H^2 \right)^2}\nonumber\\
    &\qquad \qquad  - \frac{\left(525+373 a^2H^2 \right)\arctan(aH)}{96 a^5 H^5} \nonumber\\
    &\qquad \qquad - \frac{35i \left[Li_2(i aH)-Li_2(-iaH)\right]}{32 a^3 H^3}\Bigg\},\\
    \mathcal{S}_5 &= Ma^5  \Bigg\{ \frac{72765 +218295 a^2H^2+219255a^4H^4+74365 a^6 H^6 + 1664 a^8 H^8}{1152 a^8 H^8 (1 + a^2 H^2)^2}  \nonumber\\
    &\qquad \qquad- \frac{\left(8085 + 10780 a^2 H^2 + 3623 a^4 H^4 \right)\arctan(aH)}{128 a^9 H^9} \nonumber\\
    &\qquad \qquad - \frac{385i \left[Li_2(i aH)-Li_2(-iaH)\right]}{64 a^5 H^5}\Bigg\},\\
    \mathcal{S}_7 &= -Ma^7  \Bigg\{ \frac{7}{49152 a^{12} H^{12} \big(1 + a^2 H^2)^2} (5521230 + 18758025 a^2 H^2 + 24083631 a^4 H^4\nonumber\\
    &\qquad \qquad  \qquad+ 13962747 a^6 H^6 + 
   3105671 a^8 H^8 - 8192 a^{10} H^{10} + 8192 a^{12} H^{12}\big)  \nonumber\\
    &\qquad \qquad- \frac{5 (2576574 + 4459455 a^2 H^2 + 2432430 a^4 H^4 + 460013 a^6 H^6)\arctan(aH)}{16384 a^{13} H^{13}} \nonumber\\
    &\qquad \qquad - \frac{225225i \left[Li_2(i aH)-Li_2(-iaH)\right]}{8192 a^7 H^7}\Bigg\},\\
    \mathcal{S}_9 &= Ma^9  \Bigg\{ \frac{1}{17203200 a^{16} H^{16} \big(1 + a^2 H^2)^2}\times\nonumber\\
    &\qquad \qquad \qquad \times (176849597925 + 683124917475 a^2 H^2 + 1032894122260 a^4 H^4\nonumber\\
    &\qquad \qquad \qquad +768754011740 a^6 H^6 + 287149590415 a^8 H^8 + 
 45062550345 a^{10} H^{10} \nonumber\\
    &\qquad \qquad \qquad+ 38535168 a^{12} H^{12} - 5505024 a^{14} H^{14} + 
 19038208 a^{16} H^{16}\big)  \nonumber\\
    &\qquad \qquad - \frac{1}{98304 a^{17} H^{17}}\big(1010569131 + 2219289072 a^2 H^2 + 1664466804 a^4 H^4 \nonumber\\
    &\qquad \qquad \qquad 
   + 512143632 a^6 H^6 + 66269153 a^8 H^8\big)\arctan(aH) \nonumber\\
    &\qquad \qquad  - \frac{969969i \left[Li_2(i aH)-Li_2(-iaH)\right]}{8192 a^9 H^9}\Bigg\},
\end{align}
\end{subequations}

\subsection{Mass multipole moments} \label{Appmass}
Here are the exact expressions of the mass multipole moments $\mathcal{M}_{2l}$ for $l = \{2, 3, 4\}$.
\begin{subequations}
\begin{align}
    \mathcal{M}_4 &= Ma^4 \Big\{\frac{7/5}{1+a^2H^2} + \frac{21}{32}\frac{e^{-2Ht}}{a^9 H^9}\times \nonumber\\
    &\qquad\qquad \times\left[ 5aH \left( 21+11a^2H^2\right)-3\left(35+30a^2H^2 +3a^4H^4\right)\arctan(aH)\right]\Big\},\\
    \mathcal{M}_6 
    &= -Ma^6 \Big\{\frac{10/7}{1+a^2H^2}-\frac{1287}{1280}\frac{e^{-2Ht}}{a^{13} H^{13}}(1-a^2H^2)\times \nonumber\\
    &\quad \quad\quad \quad \quad\times \big[-7aH \left( 165+170a^2H^2+33a^4H^4\right) \nonumber\\
    &\quad \quad \quad \quad\quad \quad \quad +5 \left(231+5a^2H^2(63+21a^2H^2+a^4H^4) \right)\arctan(aH)\big] \Big\},\\
    \mathcal{M}_8 
    &= Ma^8 \Bigg\{\frac{13/9}{1+a^2H^2}-\frac{2431}{28672}\frac{e^{-2Ht}}{a^{17} H^{17}}(1-a^2H^2+a^4H^4)\times \nonumber\\
    &\quad\times \bigg[-a H (225225 + 345345 a^2 H^2 + 147455 a^4 H^4 + 15159 a^6 H^6) \nonumber\\
    &\qquad+ 
 35 (6435 + 12012 a^2 H^2 + 6930 a^4 H^4 + 1260 a^6 H^6 + 35 a^8 H^8)\arctan(aH)\bigg] \Bigg\}.
\end{align}
\end{subequations}

\bibliography{Multipole_deSitter}

\providecommand{\href}[2]{#2}\begingroup\raggedright\begin{thebibliography}{10}

\bibitem{Abbott:2020khf}
{\bf LIGO Scientific, Virgo} Collaboration, R.~Abbott {\em et.~al.}, {\it
  {GW190814: Gravitational Waves from the Coalescence of a 23 Solar Mass Black
  Hole with a 2.6 Solar Mass Compact Object}},  Astrophys. J. Lett. {\bf 896}
  (2020), no.~2 L44 [\href{http://arXiv.org/abs/2006.12611}{{\tt 2006.12611}}].

\bibitem{Abbott:2020uma}
{\bf LIGO Scientific, Virgo} Collaboration, B.~Abbott {\em et.~al.}, {\it
  {GW190425: Observation of a Compact Binary Coalescence with Total Mass $\sim
  3.4 M_{\odot}$}},  Astrophys. J. Lett. {\bf 892} (2020), no.~1 L3
  [\href{http://arXiv.org/abs/2001.01761}{{\tt 2001.01761}}].

\bibitem{LIGOScientific:2018mvr}
{\bf LIGO Scientific, Virgo} Collaboration, B.~Abbott {\em et.~al.}, {\it
  {GWTC-1: A Gravitational-Wave Transient Catalog of Compact Binary Mergers
  Observed by LIGO and Virgo during the First and Second Observing Runs}},
  Phys. Rev. X {\bf 9} (2019), no.~3 031040
  [\href{http://arXiv.org/abs/1811.12907}{{\tt 1811.12907}}].

\bibitem{Abbott:2017gyy}
{\bf LIGO Scientific, Virgo} Collaboration, B.~P. Abbott {\em et.~al.}, {\it
  {GW170608: Observation of a 19-solar-mass Binary Black Hole Coalescence}},
  Astrophys. J. {\bf 851} (2017), no.~2 L35
  [\href{http://arXiv.org/abs/1711.05578}{{\tt 1711.05578}}].

\bibitem{TheLIGOScientific:2017qsa}
{\bf LIGO Scientific, Virgo} Collaboration, B.~Abbott {\em et.~al.}, {\it
  {GW170817: Observation of Gravitational Waves from a Binary Neutron Star
  Inspiral}},  Phys. Rev. Lett. {\bf 119} (2017), no.~16 161101
  [\href{http://arXiv.org/abs/1710.05832}{{\tt 1710.05832}}].

\bibitem{Abbott:2016blz}
{\bf LIGO Scientific, Virgo} Collaboration, B.~Abbott {\em et.~al.}, {\it
  {Observation of Gravitational Waves from a Binary Black Hole Merger}},  Phys.
  Rev. Lett. {\bf 116} (2016), no.~6 061102
  [\href{http://arXiv.org/abs/1602.03837}{{\tt 1602.03837}}].

\bibitem{TheLIGOScientific:2016pea}
{\bf LIGO Scientific, Virgo} Collaboration, B.~Abbott {\em et.~al.}, {\it
  {Binary Black Hole Mergers in the first Advanced LIGO Observing Run}},  Phys.
  Rev. X {\bf 6} (2016), no.~4 041015
  [\href{http://arXiv.org/abs/1606.04856}{{\tt 1606.04856}}]. [Erratum:
  Phys.Rev.X 8, 039903 (2018)].

\bibitem{Ryan:1995wh}
F.~Ryan, {\it {Gravitational waves from the inspiral of a compact object into a
  massive, axisymmetric body with arbitrary multipole moments}},  Phys. Rev. D
  {\bf 52} (1995) 5707--5718.

\bibitem{Ryan:1997hg}
F.~D. Ryan, {\it {Accuracy of estimating the multipole moments of a massive
  body from the gravitational waves of a binary inspiral}},  Phys. Rev. D {\bf
  56} (1997) 1845--1855.

\bibitem{Berti_2015}
E.~Berti, E.~Barausse, V.~Cardoso, L.~Gualtieri, P.~Pani, U.~Sperhake, L.~C.
  Stein, N.~Wex, K.~Yagi, T.~Baker and et~al., {\it Testing general relativity
  with present and future astrophysical observations},  Classical and Quantum
  Gravity {\bf 32} (Dec, 2015) 243001.

\bibitem{Cardoso:2016ryw}
V.~Cardoso and L.~Gualtieri, {\it {Testing the black hole `no-hair'
  hypothesis}},  Class. Quant. Grav. {\bf 33} (2016), no.~17 174001
  [\href{http://arXiv.org/abs/1607.03133}{{\tt 1607.03133}}].

\bibitem{Barack:2018yly}
L.~Barack {\em et.~al.}, {\it {Black holes, gravitational waves and fundamental
  physics: a roadmap}},  Class. Quant. Grav. {\bf 36} (2019), no.~14 143001
  [\href{http://arXiv.org/abs/1806.05195}{{\tt 1806.05195}}].

\bibitem{Barausse:2020rsu}
E.~Barausse {\em et.~al.}, {\it {Prospects for Fundamental Physics with LISA}},
   Gen. Rel. Grav. {\bf 52} (2020), no.~8 81
  [\href{http://arXiv.org/abs/2001.09793}{{\tt 2001.09793}}].

\bibitem{Maggio:2021ans}
E.~Maggio, P.~Pani and G.~Raposo, {\it {Testing the nature of dark compact
  objects with gravitational waves}},
  \href{http://arXiv.org/abs/2105.06410}{{\tt 2105.06410}}.

\bibitem{Barack_2007}
L.~Barack and C.~Cutler, {\it Using lisa extreme-mass-ratio inspiral sources to
  test off-kerr deviations in the geometry of massive black holes},  Physical
  Review D {\bf 75} (Feb, 2007).

\bibitem{Babak_2017}
S.~Babak, J.~Gair, A.~Sesana, E.~Barausse, C.~F. Sopuerta, C.~P. Berry,
  E.~Berti, P.~Amaro-Seoane, A.~Petiteau and A.~Klein, {\it Science with the
  space-based interferometer lisa. v. extreme mass-ratio inspirals},  Physical
  Review D {\bf 95} (May, 2017).

\bibitem{Kastha_2019}
S.~Kastha, A.~Gupta, K.~Arun, B.~Sathyaprakash and C.~Van Den~Broeck, {\it
  Testing the multipole structure and conservative dynamics of compact binaries
  using gravitational wave observations: The spinning case},  Physical Review D
  {\bf 100} (Aug, 2019).

\bibitem{doi:10.1098}
L.~Blanchet, T.~Damour and B.~Carter, {\it Radiative gravitational fields in
  general relativity i. general structure of the field outside the source},
  Philosophical Transactions of the Royal Society of London. Series A,
  Mathematical and Physical Sciences {\bf 320} (1986), no.~1555 379--430.

\bibitem{poisson_will_2014}
E.~Poisson and C.~M. Will, {\em Gravity: Newtonian, Post-Newtonian,
  Relativistic}.
\newblock Cambridge University Press, 2014.

\bibitem{Blanchet_2014}
L.~Blanchet, {\it Gravitational radiation from post-newtonian sources and
  inspiralling compact binaries},  Living Reviews in Relativity {\bf 17} (Feb,
  2014).

\bibitem{Geroch:1970cc}
R.~P. Geroch, {\it {Multipole moments. I. Flat space}},  J. Math. Phys. {\bf
  11} (1970) 1955--1961.

\bibitem{Geroch:1970cd}
R.~P. Geroch, {\it {Multipole moments. II. Curved space}},  J. Math. Phys. {\bf
  11} (1970) 2580--2588.

\bibitem{Hansen:1974zz}
R.~O. Hansen, {\it {Multipole moments of stationary space-times}},  J. Math.
  Phys. {\bf 15} (1974) 46--52.

\bibitem{BeigSimon1}
R.~Beig and W.~Simon, {\it {Proof of a multipole conjecture due to Geroch}},
  Communications in Mathematical Physics {\bf 78} (1980), no.~1 75 -- 82.

\bibitem{BeigSimon2}
R.~Beig, W.~Simon and R.~Penrose, {\it On the multipole expansion for
  stationary space-times},  Proceedings of the Royal Society of London. A.
  Mathematical and Physical Sciences {\bf 376} (1981), no.~1765 333--341.

\bibitem{Kundu1}
P.~Kundu, {\it Multipole expansion of stationary asymptotically flat vacuum
  metrics in general relativity},  Journal of Mathematical Physics {\bf 22}
  (1981), no.~6 1236--1242.

\bibitem{Kundu2}
P.~Kundu, {\it On the analyticity of stationary gravitational fields at spatial
  infinity},  Journal of Mathematical Physics {\bf 22} (1981), no.~9
  2006--2011.

\bibitem{Beig:1981zz}
R.~Beig, {\it {The multipole expansion in general relativity}},  Acta Physica
  Austriaca {\bf 53} (1981) 249--270.

\bibitem{RevModPhys.52.299}
K.~S. Thorne, {\it Multipole expansions of gravitational radiation},  Rev. Mod.
  Phys. {\bf 52} (Apr, 1980) 299--339.

\bibitem{1983GReGr..15..737G}
Y.~{G{\"u}rsel}, {\it {Multipole moments for stationary systems: The
  equivalence of the Geroch-Hansen formulation and the Thorne formulation}},
  General Relativity and Gravitation {\bf 15} (Aug., 1983) 737--754.

\bibitem{Hoenselaers_1990}
C.~Hoenselaers and Z.~Perjes, {\it Multipole moments of axisymmetric
  electrovacuum spacetimes},  Classical and Quantum Gravity {\bf 7} (oct, 1990)
  1819--1825.

\bibitem{Sotiriou_2004}
T.~P. Sotiriou and T.~A. Apostolatos, {\it Corrections and comments on the
  multipole moments of axisymmetric electrovacuum spacetimes},  Classical and
  Quantum Gravity {\bf 21} (Nov, 2004) 5727–5733.

\bibitem{Fodor}
G.~Fodor, C.~Hoenselaers and Z.~Perjés, {\it Multipole moments of axisymmetric
  systems in relativity},  Journal of Mathematical Physics {\bf 30} (1989),
  no.~10 2252--2257.

\bibitem{Backdahl:2005be}
T.~B\"{a}ckdahl and M.~Herberthson, {\it {Explicit multipole moments of
  stationary axisymmetric spacetimes}},  Class. Quant. Grav. {\bf 22} (2005)
  3585--3594 [\href{http://arXiv.org/abs/gr-qc/0506086}{{\tt gr-qc/0506086}}].

\bibitem{B_ckdahl_2005}
T.~B\"{a}ckdahl and M.~Herberthson, {\it Static axisymmetric spacetimes with
  prescribed multipole moments},  Classical and Quantum Gravity {\bf 22} (Apr,
  2005) 1607–1621.

\bibitem{Backdahl:2006ed}
T.~B{\"a}ckdahl, {\it {Axisymmetric stationary solutions with arbitrary
  multipole moments}},  Class. Quant. Grav. {\bf 24} (2007) 2205--2215
  [\href{http://arXiv.org/abs/gr-qc/0612043}{{\tt gr-qc/0612043}}].

\bibitem{Compere:2017wrj}
G.~Comp\`ere, R.~Oliveri and A.~Seraj, {\it {Gravitational multipole moments
  from Noether charges}},  JHEP {\bf 05} (2018) 054
  [\href{http://arXiv.org/abs/1711.08806}{{\tt 1711.08806}}].

\bibitem{Bena:2020see}
I.~Bena and D.~R. Mayerson, {\it {Multipole Ratios: A New Window into Black
  Holes}},  Phys. Rev. Lett. {\bf 125} (2020), no.~22 22
  [\href{http://arXiv.org/abs/2006.10750}{{\tt 2006.10750}}].

\bibitem{Bena:2020uup}
I.~Bena and D.~R. Mayerson, {\it {Black Holes Lessons from Multipole Ratios}},
  JHEP {\bf 03} (2021) 114 [\href{http://arXiv.org/abs/2007.09152}{{\tt
  2007.09152}}].

\bibitem{Bianchi_2020}
M.~Bianchi, D.~Consoli, A.~Grillo, J.~F. Morales, P.~Pani and G.~Raposo, {\it
  Distinguishing fuzzballs from black holes through their multipolar
  structure},  Physical Review Letters {\bf 125} (Nov, 2020).

\bibitem{Bianchi_2021}
M.~Bianchi, D.~Consoli, A.~Grillo, J.~F. Morales, P.~Pani and G.~Raposo, {\it
  The multipolar structure of fuzzballs},  Journal of High Energy Physics {\bf
  2021} (Jan, 2021).

\bibitem{Mukherjee_2020}
S.~Mukherjee and S.~Chakraborty, {\it Multipole moments of compact objects with
  nut charge: Theoretical and observational implications},  Physical Review D
  {\bf 102} (Dec, 2020).

\bibitem{Seraj_2017}
A.~Seraj, {\it Multipole charge conservation and implications on
  electromagnetic radiation},  Journal of High Energy Physics {\bf 2017} (Jun,
  2017).

\bibitem{Abbott:1981ff}
L.~F. Abbott and S.~Deser, {\it {Stability of Gravity with a Cosmological
  Constant}},  Nucl. Phys. B {\bf 195} (1982) 76--96.

\bibitem{Barnich:2001jy}
G.~Barnich and F.~Brandt, {\it {Covariant theory of asymptotic symmetries,
  conservation laws and central charges}},  Nucl. Phys. B {\bf 633} (2002)
  3--82 [\href{http://arXiv.org/abs/hep-th/0111246}{{\tt hep-th/0111246}}].

\bibitem{ABKIII}
A.~Ashtekar, B.~Bonga and A.~Kesavan, {\it {Asymptotics with a positive
  cosmological constant: III. The quadrupole formula}},  Phys. Rev. D {\bf 92}
  (2015), no.~10 104032 [\href{http://arXiv.org/abs/1510.05593}{{\tt
  1510.05593}}].

\bibitem{Rajeev:2019okd}
K.~Rajeev, S.~Chakraborty and T.~Padmanabhan, {\it {Generalized Schwinger
  effect and particle production in an expanding universe}},  Phys. Rev. D {\bf
  100} (2019), no.~4 045019 [\href{http://arXiv.org/abs/1904.03207}{{\tt
  1904.03207}}].

\bibitem{ABKI}
A.~Ashtekar, B.~Bonga and A.~Kesavan, {\it {Asymptotics with a positive
  cosmological constant: I. Basic framework}},  Class. Quant. Grav. {\bf 32}
  (2015), no.~2 025004 [\href{http://arXiv.org/abs/1409.3816}{{\tt
  1409.3816}}].

\bibitem{deVega:1998ia}
H.~de~Vega, J.~Ramirez and N.~G. Sanchez, {\it {Generation of gravitational
  waves by generic sources in de Sitter space-time}},  Phys. Rev. D {\bf 60}
  (1999) 044007 [\href{http://arXiv.org/abs/astro-ph/9812465}{{\tt
  astro-ph/9812465}}].

\bibitem{DHI}
G.~Date and S.~J. Hoque, {\it Gravitational waves from compact sources in a de
  sitter background},  Phys. Rev. D {\bf 94} (Sep, 2016) 064039.

\bibitem{ABKII}
A.~Ashtekar, B.~Bonga and A.~Kesavan, {\it {Asymptotics with a positive
  cosmological constant. II. Linear fields on de Sitter spacetime}},  Phys.
  Rev. D {\bf 92} (2015), no.~4 044011
  [\href{http://arXiv.org/abs/1506.06152}{{\tt 1506.06152}}].

\bibitem{Carter:1968ks}
B.~Carter, {\it {Hamilton-Jacobi and Schrodinger separable solutions of
  Einstein's equations}},  Commun. Math. Phys. {\bf 10} (1968), no.~4 280--310.

\bibitem{GIBBONS200549}
G.~Gibbons, H.~Lü, D.~N. Page and C.~Pope, {\it The general kerr–de sitter
  metrics in all dimensions},  Journal of Geometry and Physics {\bf 53} (2005),
  no.~1 49--73.

\bibitem{Olea:2005gb}
R.~Olea, {\it {Mass, angular momentum and thermodynamics in four-dimensional
  Kerr-AdS black holes}},  JHEP {\bf 06} (2005) 023
  [\href{http://arXiv.org/abs/hep-th/0504233}{{\tt hep-th/0504233}}].

\bibitem{Akcay:2010vt}
S.~Akcay and R.~A. Matzner, {\it {Kerr-de Sitter Universe}},  Class. Quant.
  Grav. {\bf 28} (2011) 085012 [\href{http://arXiv.org/abs/1011.0479}{{\tt
  1011.0479}}].

\bibitem{Chrusciel:2015sna}
P.~T. Chru\'sciel, J.~Jezierski and J.~Kijowski, {\it {Hamiltonian dynamics in
  the space of asymptotically Kerr\textendash{}de Sitter spacetimes}},  Phys.
  Rev. D {\bf 92} (2015), no.~8 084030
  [\href{http://arXiv.org/abs/1507.03868}{{\tt 1507.03868}}].

\bibitem{Hinterbichler_2014}
K.~Hinterbichler, L.~Hui and J.~Khoury, {\it An infinite set of ward identities
  for adiabatic modes in cosmology},  Journal of Cosmology and Astroparticle
  Physics {\bf 2014} (Jan, 2014) 039–039.

\bibitem{mirbabayi2016weinberg}
M.~Mirbabayi and M.~Simonović, {\it Weinberg soft theorems from weinberg
  adiabatic modes},  \href{http://arXiv.org/abs/1602.05196}{{\tt 1602.05196}}.

\bibitem{Hamada_2018}
Y.~Hamada and G.~Shiu, {\it Infinite set of soft theorems in gauge-gravity
  theories as ward-takahashi identities},  Physical Review Letters {\bf 120}
  (May, 2018).

\end{thebibliography}\endgroup

\end{document}